%% file: main.tex
\documentclass[sn-mathphys-num]{sn-jnl}

\usepackage{graphicx}%
\usepackage{multirow}%
\usepackage{amsmath,amssymb,amsfonts}%
\usepackage{amsthm}%
\usepackage{mathrsfs}%
\usepackage[title]{appendix}%
\usepackage{xcolor}%
\usepackage{textcomp}%
\usepackage{manyfoot}%
\usepackage{booktabs}%
\usepackage{algorithm}%
\usepackage{listings}%
\usepackage{subfiles}
\usepackage{bm}
\usepackage{tabularx}
\usepackage{algorithmic}
\usepackage{lscape}

\theoremstyle{thmstyleone}%
\theoremstyle{thmstyletwo}%

\theoremstyle{thmstylethree}%

\begin{document}

\title[Spin-Glass-Based Traffic Optimization for Urban Networks]{Simulated Bifurcation with High-dimensional Expansion for Traffic Signal Optimization on Real-world Networks}


\author[1,2]{\fnm{Shengda} \sur{Zhao}}\email{24110511@bjtu.edu.cn}

\author[3]{\fnm{Zhekun} \sur{Liu}}\email{24120359@bjtu.edu.cn}

\author[1,2]{\fnm{Jiaxin} \sur{Yu}}\email{24110506@bjtu.edu.cn}

\author[4]{\fnm{Bocheng} \sur{Ju}}\email{24115070@bjtu.edu.cn}

\author[5]{\fnm{Liang} \sur{Wang}}\email{wljean@126.com}

\author*[5]{\fnm{Xiaodong} \sur{Zhang}}\email{zhangxd-bicp@outlook.com}

\author*[1,2]{\fnm{Xinghua} \sur{Zhang}}\email{zhangxh@bjtu.edu.cn}

\affil*[1]{\orgdiv{School of Physical Sciences and Engineering}, \orgname{Beijing Jiaotong University}, \orgaddress{\street{Haidian District}, \city{Beijing}, \postcode{100044}, \country{China}}}

\affil[2]{\orgdiv{Beijing Key Laboratory of Novel Materials Genetic Engineering and Application for Rail Transit}, \orgname{Beijing Jiaotong University}, \orgaddress{\street{Haidian District}, \city{Beijing}, \postcode{100044}, \country{China}}}

\affil[3]{\orgdiv{School of Computer Science and Technology}, \orgname{Beijing Jiaotong University}, \orgaddress{\street{Haidian District}, \city{Beijing}, \postcode{100044}, \country{China}}}

\affil[4]{\orgdiv{National School of Elite Engineers}, \orgname{Beijing Jiaotong University}, \orgaddress{\street{Haidian District}, \city{Beijing}, \postcode{100044}, \country{China}}}

\affil[5]{\orgdiv{}, \orgname{Beijing Municipal Institute of City Planning \& Design}, \orgaddress{\street{Xicheng District}, \city{Beijing}, \postcode{100045}, \country{China}}}


\abstract{
    In the context of accelerating urbanization and increasingly severe traffic congestion, optimizing traffic signal systems to enhance road throughput and reduce congestion has become a critical issue that requires attention. 
    This study proposes a short-term traffic situation prediction model based on real-world road topologies and a typical four-way, eight-phase traffic signal control scheme. The model takes into account the traffic flow disparities across different directions and the frequency of signal phase changes, integrating these factors into an optimization objective function aimed at achieving global optimization of urban traffic. The structure of this objective function is analogous to the spin-glass systems in statistical physics.

    For this model, a Simulated Bifurcation optimization algorithm is introduced, with traditional simulated annealing optimization methods used as a benchmark. The results demonstrate that the Simulated Bifurcation algorithm outperforms simulated annealing in terms of both efficiency and effectiveness.
    Using real traffic flow and road network data from Beijing, we initialized the model and conducted numerical optimization experiments. The experimental results indicate that the Simulated Bifurcation algorithm significantly outperforms simulated annealing in terms of computational efficiency, effectively addressing combinatorial optimization problems in systems with multiple spin interactions, and reducing the time complexity of the problem to $O(N^{1.35})$. This successful solution addresses the NP-hard problem of global traffic signal optimization.
    
    More importantly, the signal phase patterns generated by the Simulated Bifurcation algorithm are fully compatible with the operational requirements of actual traffic signal systems, showcasing its enormous potential and applicability in optimizing signal control in real-world traffic networks, especially for large, complex urban traffic systems. 
    This work provides strong theoretical support and practical foundations for the future implementation of urban traffic management and intelligent transportation systems.
}

\keywords{Traffic Signal Optimization, Simulated Bifurcation Algorithm,Combinatorial Optimization, Spin-Glass Systems}



\maketitle


\subfile{Sections/Introduction.tex}

\section{Results}
    \subfile{Sections/Results.tex}

\section{Discussion}
    \subfile{Sections/Discussion.tex}

\section{Methods}
    \subfile{Sections/Methods.tex}

\section{Data and Code availability}
    The data and code used in this study are publicly available on GitHub. The repository includes all necessary scripts and datasets to reproduce the results presented in this paper. You can access the repository at the following link: [https://github.com/liftes/TrafficSignal.git].


\bibliography{main}

\section{Acknowledgements}
    This work was supported by the Fundamental Research Funds for the Central Universities 2023JBWZD002.

\section{Author contributions}
    S.Z. wrote the manuscript and developed all the computational code. Z.L. and B.J. conducted the preliminary exploration of the SB algorithm, with Z.L. also assisting in the writing of supporting documents. J.Y. contributed to deriving the theory of the SB algorithm. L.W. and X.Z. provided the data support. All work was completed under the guidance of X.Z. 

\section{SS1: Combinatorial Optimization Problems}~
    \subfile{SupplementarySections/SS1.tex}

\section{SS2: Real-state Encoding of Intersection States}
    \subfile{SupplementarySections/SS2.tex}

\section{SS3: Cost Function: Hamiltonian Represented in Spin Glass Form}
    \subfile{SupplementarySections/SS3.tex}

\section{SS4: Extension of the SB Algorithm}
    \subfile{SupplementarySections/SS4.tex}

\end{document}

%% file: Sections/Introduction.tex
As the global urbanization process accelerates, the pressure on both intra-city and inter-city transportation continues to rise, making the development of intelligent transportation systems one of the key research topics of the 21st century \cite{E.A.Stanciu25Optimization, L.Butler2020Smart, Zhu2020Parallel}. 
As a crucial component of urban transportation, road traffic accounts for more than 50\% of the transportation tasks \cite{Vitkunas2021Assessment}. However, with the rapid increase in the number of vehicles, traffic congestion on urban roads at all levels has become increasingly severe, leading to serious secondary issues such as air pollution. 
In response to these challenges, alongside the improvement of public transportation infrastructure, designing more efficient traffic signal control systems has become a central focus of academic research \cite{Bharadiya2023Artificial, Nigam2023Review, Samaei2024Using}.

In recent years, research on intelligent traffic signal control has primarily focused on optimizing cycle times and red/green splits to better balance traffic flow across different roads \cite{Hao2022Convergence,Yu2018Integrated}. Traditional methods, such as the Webster model based on queuing theory, provide effective calculations for signal cycle times at single intersections and remain widely used due to their practicality \cite{Webster1958Traffic}.
To address dynamic traffic flows, adaptive traffic signal control systems, such as SCATS and SCOOT, leverage real-time traffic data to dynamically adjust signal phases and durations, enabling the system to adapt to complex and changing traffic conditions \cite{Lowrie1990Scats, Hunt1981SCOOTaTR}. 
With advancements in artificial intelligence, reinforcement learning algorithms have been introduced into traffic signal control, allowing systems to iteratively learn and optimize control strategies through interaction with their environment. 
Additionally, the rapid development of Vehicle-to-Infrastructure (V2I) technology has enabled communication between vehicles and infrastructure, where precise vehicle position and speed information further enhance the optimization of traffic signal control \cite{Cabrejas-Egea2021Reinforcement, Du2023Safelight}. 
Meanwhile, fuzzy logic control provides a flexible solution for dynamically adjusting signal phases and durations by addressing uncertainties inherent in traffic data through fuzzification \cite{Bi2014Type2, Nair2007fuzzy, Zhang2017Application}. 
The integration of these technologies not only advances the intelligence level of urban traffic signal control but also offers new possibilities for mitigating increasingly severe traffic congestion.

Although research on optimizing traffic signals at single intersections has been widely applied, studies have shown that single-point optimization can only address vehicle delays at individual intersections or in specific directions, but it is insufficient to effectively alleviate traffic congestion on a citywide scale. 
This limitation arises because traffic flow is often interconnected across multiple intersections within the same area, where localized optimization at individual intersections may inadvertently lead to systemic congestion \cite{Yu2018Optimal}. Moreover, urban traffic patterns exhibit significant dynamism due to factors such as weather conditions, differences between weekdays and holidays, and variations across time periods. As a result, traffic signals need to adjust parameters like cycle times and red/green splits in real time based on the current traffic situation.
In this context, the core objective of improving the overall efficiency of urban traffic systems is to consider the interdependencies of traffic flows across intersections and achieve global optimization. To address the complexity of urban traffic systems, an increasing number of studies are focusing on optimizing traffic signal control for entire urban road networks \cite{Qadri2020Stateofart}. 
However, given the large scale of urban traffic networks, particularly in megacities where the number of intersections can reach $10^3$ to $10^4$, achieving such large-scale global optimization requires precise coordination and integrated scheduling among traffic signals at numerous intersections. 
The control of traffic signals at each intersection not only involves multiple phase transitions but also demands coordination across intersections, causing the scale and complexity of the optimization problem to grow rapidly. This multi-phase, integer-variable, large-scale optimization problem poses significant challenges for algorithm design.

The global optimization of traffic signals can be reduced to a classic combinatorial optimization problem, whose complexity belongs to the NP-hard problem \cite{PAIK1995NETWORK, Renfrew2009Traffic, Tchuitcheu2020Internet}. 
The objectives of traffic signal control, such as minimizing average waiting times and maximizing traffic throughput, require finding a global optimum under finite states and multiple constraints. As the number of intersections increases, the scale and complexity of the problem grow exponentially.
To address the computational bottlenecks of NP-hard problems, several novel optimization methods have emerged in recent years. For instance, multi-agent systems (MAS) treat each intersection as an independent agent and achieve global traffic optimization through coordination and communication among agents, offering a new approach for intelligent traffic management in urban areas \cite{Chu2019Multiagent, El-Tantawy2010agentbased}.
Some heuristic optimization algorithms have shown promising results in small-scale networks. For example, Ge et al. applied simulated annealing \cite{Ge2014EnergySustainable}, while Renfrew et al. utilized the Ant Colony Optimization (ACO) algorithm \cite{Renfrew2012Traffic}. However, as the network scale expands, the computational time required by numerical algorithms increases dramatically, limiting their applicability to large-scale traffic systems \cite{Qadri2020Stateofart, Eom2020traffic}.

Designing effective solvers to overcome the challenges of NP-hard combinatorial optimization problems is a major focus in current research. Recently, the work by Inoue et al. provided a novel approach to this problem by constructing a mathematical model that links traffic signals with traffic flow in grid-based networks. Their study successfully reformulated the problem into an Ising model, which can be efficiently solved using physical methods \cite{DaisukeInoue2021Traffic}.
This research introduced an innovative perspective by establishing a connection between traffic signal optimization and Ising solvers, equating the traffic signal optimization problem to a physical problem. This approach offers a fresh angle for addressing combinatorial optimization challenges in traffic systems. 
However, the classical Ising model has limitations in practical applications. It is inherently designed for simple two-phase traffic signal systems (red and green lights), whereas real-world scenarios often involve multi-phase traffic signals that account for left-turn directions at intersections. Moreover, the assumption of a grid-based network restricts the model's applicability to the highly dynamic and complex traffic networks observed in real-world urban environments.
In particular, when considering the intricate spatial topology of road networks and the interdependencies between intersections, grid-based models fail to effectively capture the characteristics of realistic traffic systems. Developing practical models capable of accurately describing complex road network structures while supporting multi-phase traffic signals, along with targeted, efficient combinatorial optimization algorithms, remains a critical challenge that demands immediate attention in the field.

To address these challenges, we redefined the dynamic relationship between traffic flow and signal phases under real-world road network topologies, based on the eight-phase traffic signal model for individual intersections. 
Specifically, we established a direct correspondence between traffic signal optimization and the physical spin-glass model by defining the variance of traffic flows across different roads as the Hamiltonian. As a robust optimization tool, the spin-glass model effectively captures the constraints and uncertainties inherent in large-scale complex systems.
To enhance the efficiency of solving the spin-glass model, we extended the Simulated Bifurcation algorithm, originally designed for the Ising model, into higher dimensions using field theory. This approach reformulates the combinatorial optimization problem into a set of dynamical equations, which are then solved using the finite difference method. This improvement significantly enhances the computational efficiency and scalability of the algorithm, reducing its time complexity to \(O(N^{1.35})\). This advancement makes the algorithm feasible for large-scale applications in real-world road networks.
Moreover, all numerical experiments were conducted using real-world road network data from Beijing, demonstrating the algorithm's practicality in real-life scenarios. The successful application of this method to the traffic signal optimization of Beijing—a megacity with one of the most complex urban road networks—provides valuable insights and scalable solutions for traffic signal optimization in other large cities.

%% file: Sections/Results.tex
\subsection{Computational Model}
    This study introduces the Spin Glass Model as an extension of the Ising model to address the complex problem of traffic flow optimization. Compared to the traditional Ising model, the Spin Glass Model offers greater flexibility, effectively overcoming the limitations of the former. It provides a more precise framework for describing the traffic signal system in transportation networks, and on this basis, solves combinatorial optimization problems (Supplementary Section 1). 
    Moreover, the model can dynamically adapt to changes in road segments, intersections, and traffic conditions, offering an efficient solution for optimizing traffic networks. By incorporating this model, this research is able to more accurately optimize traffic signal control systems with multiple phases, directions, and dynamic variations. This significantly enhances the overall efficiency and reliability of urban transportation systems, providing strong theoretical support and practical guidance for future traffic management and optimization.

    \subsubsection{Real-state encoding of intersection states}

        This study is based on traditional signal control models, selecting four-way intersections and eight-phase intersections as the fundamental units for modeling urban road networks. 
        The eight-phase intersection model is currently the most widely used intersection model and effectively captures the complexity of urban traffic.
        The research uses graphs to describe the topological structure of road connections.
        This not only simplifies the representation of the road network but also provides a solid foundation for subsequent analysis and optimization.

        \begin{figure}[tp]
            \centering
            \includegraphics[width=\linewidth]{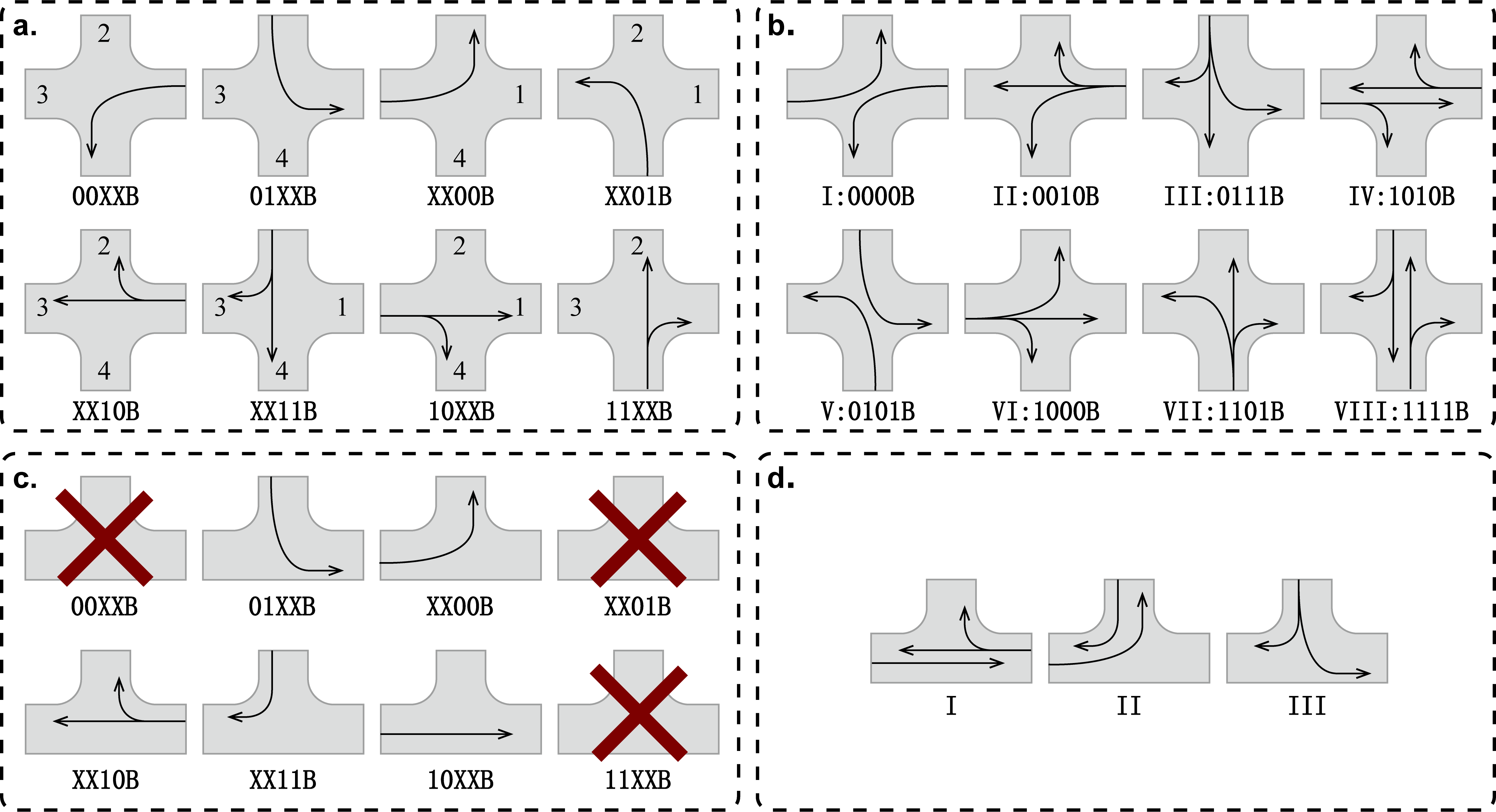}
            \caption{
                (a) Basic steering direction encoding for a four-way intersection with eight phases;
                (b) Real-state encoding of a cross intersection using the four-way intersection eight-phase encoding;
                (c) Basic steering direction encoding for a T-junction compatible with the four-way intersection eight-phase model;
                (d) Real-state encoding for a T-junction.
            }
            \label{fig:TrafficPhaseCode}
        \end{figure}

        Fig. \ref{fig:TrafficPhaseCode} illustrates the basic steering direction encoding method we employed.
        The traffic signals of an eight-phase intersection can be viewed as a combination of the basic steering direction patterns for each road direction. Since most right turns and straight movements at cross-intersections can occur simultaneously, we define each direction's left turn and straight movement combined with the right turn as a basic steering direction pattern.
        As shown in Fig. \ref{fig:TrafficPhaseCode}(a), this results in eight basic steering direction patterns, which serve as the foundation for the modeling.

        To describe the relationships between different signal phases, we encoded the basic steering direction patterns using a two-bit binary number and a two-bit mask (XX). 
        The advantage of this approach lies in its direct connection between traffic flow variations and binary digits, providing a clear mathematical framework for flow evolution. 
        This encoding method allows for efficient representation of traffic flow characteristics for similar turning phases, leading to a more precise description and optimization of traffic signal transitions.        
        Additionally, this method ensures that the barriers between adjacent passing modes are low. For example, transitioning from 1101B to 1111B only requires overcoming a one-bit binary difference. 
        This characteristic facilitates smooth transitions in the optimal state search, enhancing the convergence and stability of numerical algorithms, and providing more accurate parameter support for traffic optimization. A more detailed explanation of the encoding method can be found in Supplementary Section 2.

    \subsubsection{Cost function: Hamiltonian represented in spin glass form}

        According to the derivation in Supplementary Section 3, the overall Hamiltonian of the system can be expressed as:
        \begin{equation}
            H[\bm{x}(t),\bm{x}(t-1)] = H_\text{q}[\bm{x}(t)] + H_\text{d}[\bm{x}(t),\bm{x}(t-1),t].
            \label{eq:H-base}
        \end{equation}
        Here, the flow deviation cost function \(H_\text{q}\) represents the variance of the instantaneous traffic flow \(q(i,j,t)\) in each direction, where \( q(i,j,t) = q_{i,j}(t) \) represents the traffic flow on the road connecting intersection \(i\) and intersection \(j\) at time \(t\).
        Generally, in large-scale urban road networks, the more evenly distributed the vehicles are, the smoother the traffic conditions tend to be. 
        Specifically, when the instantaneous traffic flow in all directions at an intersection becomes consistent, it signifies higher efficiency at the intersection, meaning \(H_\text{q} = 0\).
        
        The signal phase change cost function \(H_{\text{d}}\) measures the cost of signal switching. 
        This function aims to smooth the traffic signal transitions and prevent frequent changes in the signal lights within short periods, thereby reducing the negative impact on traffic flow stability.
        
        Due to the complexity of real road networks, further mathematical simplification and regularization of the model are not performed here. 
        However, it is noteworthy that for the variable \( \bm{x} \), since \( q_{i,j} \) is a summation of quadratic terms of \( \bm{x} \), it forms a quadratic homogeneous expression. 
        As a result, the Hamiltonian \( H_\text{q} \) contains fourth-order terms in \( \bm{x} \), similar to the multi-spin interaction terms in the Spin Glass Model.        
        On the other hand, the \( H_{\text{d}} \) term reflects the energy change caused by the variation in the state \( \bm{x} \), and is represented by a summation of quadratic terms of \( \bm{x} \), analogous to the second-order auto-correlation terms in the Spin Glass Model. 
        Therefore, the model form represented by equation (\ref{eq:H-base}) is very similar to the four-spin interaction model.

\subsection{Solution Algorithm}

    By applying theoretical models from statistical physics to the global optimization problem of traffic signals, we are able to leverage existing physical optimization algorithms to solve combinatorial optimization problems, or construct corresponding physical machines to replace classical computers. 
    This approach avoids the computational bottlenecks faced by traditional algorithms under the Von Neumann architecture when solving NP-hard problems \cite{chang2024quantum, Inagaki2016coherent, Mohseni2022Ising}. 
    The solution time for NP-hard problems grows exponentially with the increase in problem size, making it extremely difficult to solve large-scale instances. 
    This issue is particularly severe in highly dynamic and heterogeneous environments such as urban traffic networks, where traditional computational methods often fail to effectively handle complex optimization problems due to their large computational burden and inefficiency. 
    In contrast, physical optimization algorithms and physical computers, such as quantum computing and simulated annealing, can provide potential solutions by simulating natural optimization processes. 
    These methods show significant advantages, especially when solving complex systems \cite{GotoHayato2021Highperformance, SaavanPatel2020Ising}.

    In physics research, solving the spin glass model with multi-spin interactions still presents significant challenges. 
    The computational complexity of the spin glass model is high, especially when the number of spins is large, as the model contains numerous random couplings and local minima, making traditional numerical methods difficult to apply \cite{auffinger2013random, Ros2023highdimensional}. 
    Thus, despite the model's rich descriptive capabilities in theory, significant technical challenges remain for its effective application to real-world problems. 
    Currently, the academic community is exploring efficient numerical methods for solving spin glass models, such as variational methods \cite{billoire2018dynamic, franz2001exact} and Monte Carlo methods \cite{kiss2024complete, mo2023nature}, while also attempting to extend numerical algorithms developed for the Ising model (e.g., simulated annealing \cite{goto2019combinatorial}, quantum annealing \cite{VBapst2013Thermal, DaisukeInoue2021Traffic}, and Boltzmann machines \cite{FrancescoDAngelo2020Learning}) to solve the spin glass model.

    This paper draws on the simulated bifurcation (SB) algorithm, which has performed excellently on the Ising model in recent years, and extends it for use with the spin glass model in combination with self-consistent field theory. 
    The SB algorithm can gradually approach the optimal solution through the branching behavior of physical systems, with a computation process that exhibits strong parallelism and adaptability, making it especially suitable for solving multi-spin systems. 
    After integrating with self-consistent field theory, the algorithm’s convergence and stability are further enhanced, showing exceptional performance when handling multi-body interactions and inhomogeneous systems. 
    The recursive equation for the extended SB algorithm can be written as:
    \begin{equation}
        \dot{x_i}^{(c)}=\frac{\partial H_\text{SB}}{\partial y_i^{(c)}},
        \label{eq:xic}
    \end{equation}
    \begin{equation}
        \dot{y_i}^{(c)}=-\frac{\partial H_\text{SB}}{\partial \tilde{x}_i^{(c)}}.
        \label{eq:yic}
    \end{equation}
    Here,
    \begin{equation}
        H_\text{SB}(\bm{\tilde{x}}, y)=
        \frac{a_0}{2}\sum_{c=1}^{4}\sum_{i=1}^{N}{(y_i^{(c)})}^2
        +V_\text{SB}(\bm{\tilde{x}}),
    \end{equation}
    \begin{equation}
        V_\text{SB}(\bm{\tilde{x}})=
        \sum_{c=1}^{4}\sum_{i=1}^{N}
        \left\{ \frac{1}{2} [a_0-a(\tau)] {(\tilde{x}_i^{(c)})}^2 \right\}
        + H(\bm{\tilde{x}}).
        \label{eq:V_SB}
    \end{equation}
    In the above equation, \(a_0\) and \(c_0\) are constants that control the strength of the harmonic potential \((\tilde{x}^{(c)}_i)^2\) and the external field \(H\) in the optimization objective, respectively. 
    Typically, \(a_0 = c_0 = 1\) is chosen. In this case, the bifurcation function \(a(\tau)\) is a linear function of the iteration step, i.e., \(a(\tau) = \tau\). 
    Here, \(\tau = \Delta \tau \times Iter\), where \(\Delta \tau\) is the iteration step size and \(Iter\) is the current iteration number, with the maximum iteration count given by \(Iter_{\text{max}} = a_0 / \Delta t\). 
    Thus, as \(\tau\) increases, \(a(\tau)\) increases linearly until \(a_0 - a(\tau) = 0\), at which point the harmonic potential term loses its constraint, and the system converges.

    It is important to note that, to ensure the validity of the final result, a Lagrangian function \(H_w\) must be introduced in equation (\ref{eq:H-base}) to limit the free evolution of the system, ensuring that the search is conducted only within the allowed state space. 
    Furthermore, to guarantee the effectiveness of the SB algorithm, \(H_w\) should be a continuous function and continuously differentiable within the domain of \(\tilde{x}^{(c)}_i\).
    Therefore, equation (\ref{eq:H-base}) must be rewritten as:
    \begin{equation}
        H[\bm{x}(t),\bm{x}(t-1)] = H_\text{q}[\bm{x}(t)] + H_\text{d}[\bm{x}(t),\bm{x}(t-1),t] + H_\text{w}[\bm{\tilde{x}}(t)],
        \label{eq:H-base-fix}
    \end{equation}
    and then substituted into equation (\ref{eq:V_SB}) to obtain \(H_{\text{SB}}\).
    Further details of the algorithm are provided in Supplementary Section 4.1.

\subsection{Numerical Experiments}

    In this study, we propose a spin glass-based global optimization model for traffic signals that can be applied to real-world data, and introduce the SB algorithm to solve this model. 
    By comparing it with the SA algorithm (Supplementary Section 4.2), we validate its performance in real urban road networks. 
    Through numerical experiments, we conduct an in-depth analysis of the algorithm's convergence, stability, and its responsiveness to changes in traffic signal phases within actual traffic networks.

    \begin{figure}[t]
        \centering
        \includegraphics[width=0.9\linewidth]{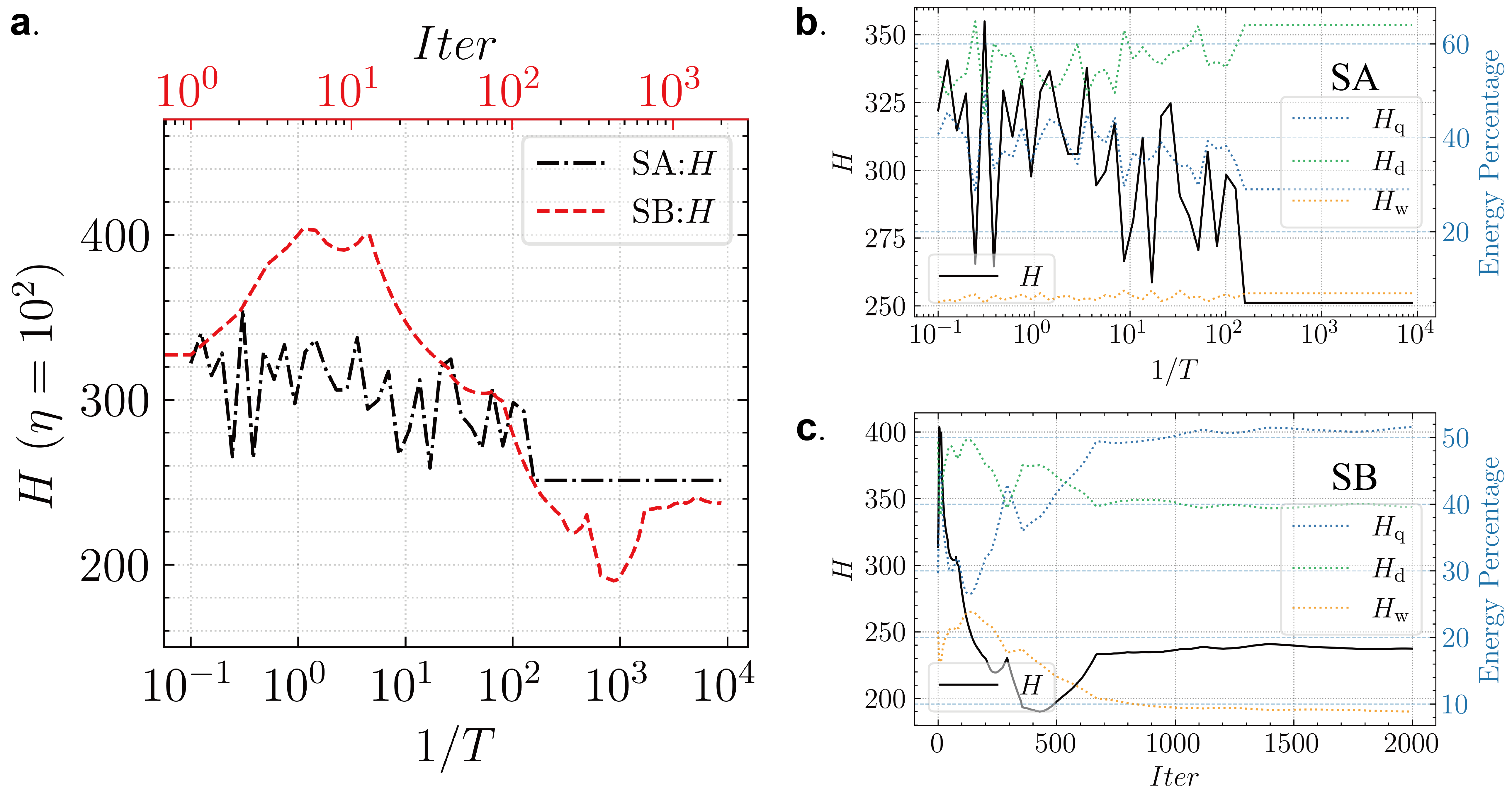}
        \caption{
            (a) Comparison of $H$ convergence for SA and SB algorithms when $\eta = 10^2$. The SA algorithm is represented by the black line, with the system converging as the temperature decreases, while the SB algorithm is represented by the red line, with the system converging as the number of iterations increases.
            (b) and (c) Show the overall convergence behavior of the SA and SB algorithms, respectively. The right-hand axis indicates the contribution percentages of the three Hamiltonians: $H_{\text{p}}$, $H_{\text{d}}$, and $H_{\text{w}}$.
        }
        \label{fig:Result-H}
    \end{figure}

    Fig. \ref{fig:Result-H}(a) shows the typical convergence behavior of the SA and SB algorithms under the condition \(\eta = 10^2\) across different time scales. 
    The SA algorithm gradually converges as the simulation temperature \(T\) decreases (i.e., as \(1/T\) increases), while the SB algorithm converges progressively with the increase in \(Iter\). 
    This indicates that both algorithms are capable of effectively searching for and finding a set of optimized states \(\bm{x}\) for the system.
    Further comparison of Fig. \ref{fig:Result-H}(b) and Fig. \ref{fig:Result-H}(c) provides a clearer picture of the differences in the Hamiltonian distribution between the two algorithms. 
    In the SA algorithm, the proportion of \(H_{\text{w}}\) remains consistently low, around 10\%. 
    This suggests that the system explores the phase states within the forbidden set \(\varLambda\) infrequently during the entire search process, indicating that its optimization is mainly confined to the state space of the allowed set \(M\).    
    In contrast, the variation of \(H_{\text{w}}\) in the SB algorithm shows an initial rise followed by a decrease, with the drop in \(H_{\text{w}}\) around \(Iter = 500\) corresponding to a brief increase in the total Hamiltonian \(H\). 
    This phenomenon suggests that the SB algorithm makes full use of the phases in the set \(\varLambda\) as transition states during the convergence process, helping the system to more smoothly move past local optima and achieve more efficient global convergence.

    \begin{figure*}[t]
        \centering
        \includegraphics[width=0.9\linewidth]{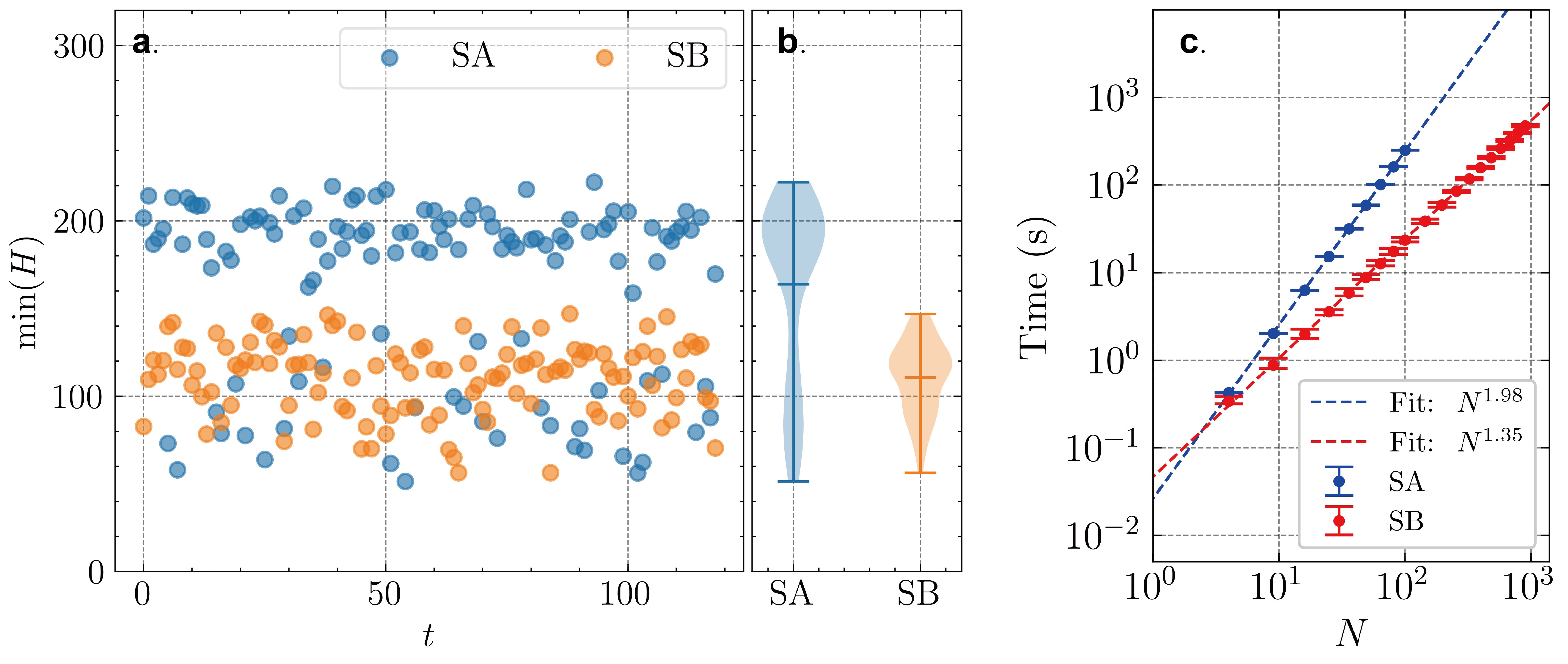}
        \caption{
            (a) The scatter plot shows the system states at 120 consecutive time steps calculated from real data, with the corresponding Hamiltonian $H$ values during convergence. 
            (b) The violin plot on the right illustrates the mean and distribution of $H$ at these time steps.
            (c) The algorithm runtime is shown for networks with different numbers of nodes, calculated using randomly generated topological networks.
        }
        \label{fig:Result-Benchmark}
    \end{figure*}

    By reading real traffic data (for detailed methods, see Methods), we selected 120 consecutive time steps from the actual road network for the global optimization test of traffic signals, with a 5-minute interval between each sampling time. 
    Fig. \ref{fig:Result-Benchmark}(a) shows the values of \(\text{min}(H)\) obtained by the SA and SB algorithms for these real data.
    It is important to note that in this test, the values of \(q_{i,j}\), left turn probability \(\alpha_{i,j}\), and right turn probability \(\beta_{i,j}\) remain consistent at each time step.
    This setting helps to effectively compare the performance differences between the SA and SB algorithms under the same conditions.
    
    Fig. \ref{fig:Result-Benchmark}(b) shows the final distribution of \(\text{min}(H)\) for both algorithms. 
    It is clearly evident that the SA algorithm exhibits a wider distribution and follows a typical bimodal distribution pattern, with only a few time steps achieving \(\text{min}(H)\) values close to those of the SB algorithm. 
    This indicates that the SA algorithm has weak capabilities to overcome the barriers between different local optima during the optimization process, limiting its global search effectiveness.
    In contrast, the distribution of \(\text{min}(H)\) for the SB algorithm is narrower, with an overall lower average value compared to the SA algorithm. 
    This indicates that the SB algorithm has stronger search capabilities and is more effective at finding optimized solutions. 
    This significant performance improvement is closely tied to the basic assumptions of the SB algorithm.     
    By introducing the continuous spin assumption and adopting a strategy of progressively reducing the depth of potential well constraints, the SB algorithm can better capture the local structures of the Hamiltonian potential surface during the search process. 
    Furthermore, this strategy allows for the smooth transition of the system from one local optimum to another in the vicinity of the energy surface. 
    In other words, the SB algorithm is able to effectively overcome the barriers between different local optima.
    As a result, the SB algorithm demonstrates a stronger global search capability during the optimization process, avoiding the pitfall of the SA algorithm becoming trapped in local optima. 
    Therefore, the SB algorithm exhibits superior performance and higher stability when handling large-scale combinatorial optimization problems, such as global traffic signal optimization in real-world scenarios, especially in complex environments with multi-spin interactions.

    \begin{figure}[ht]
        \centering
        \includegraphics[width=\linewidth]{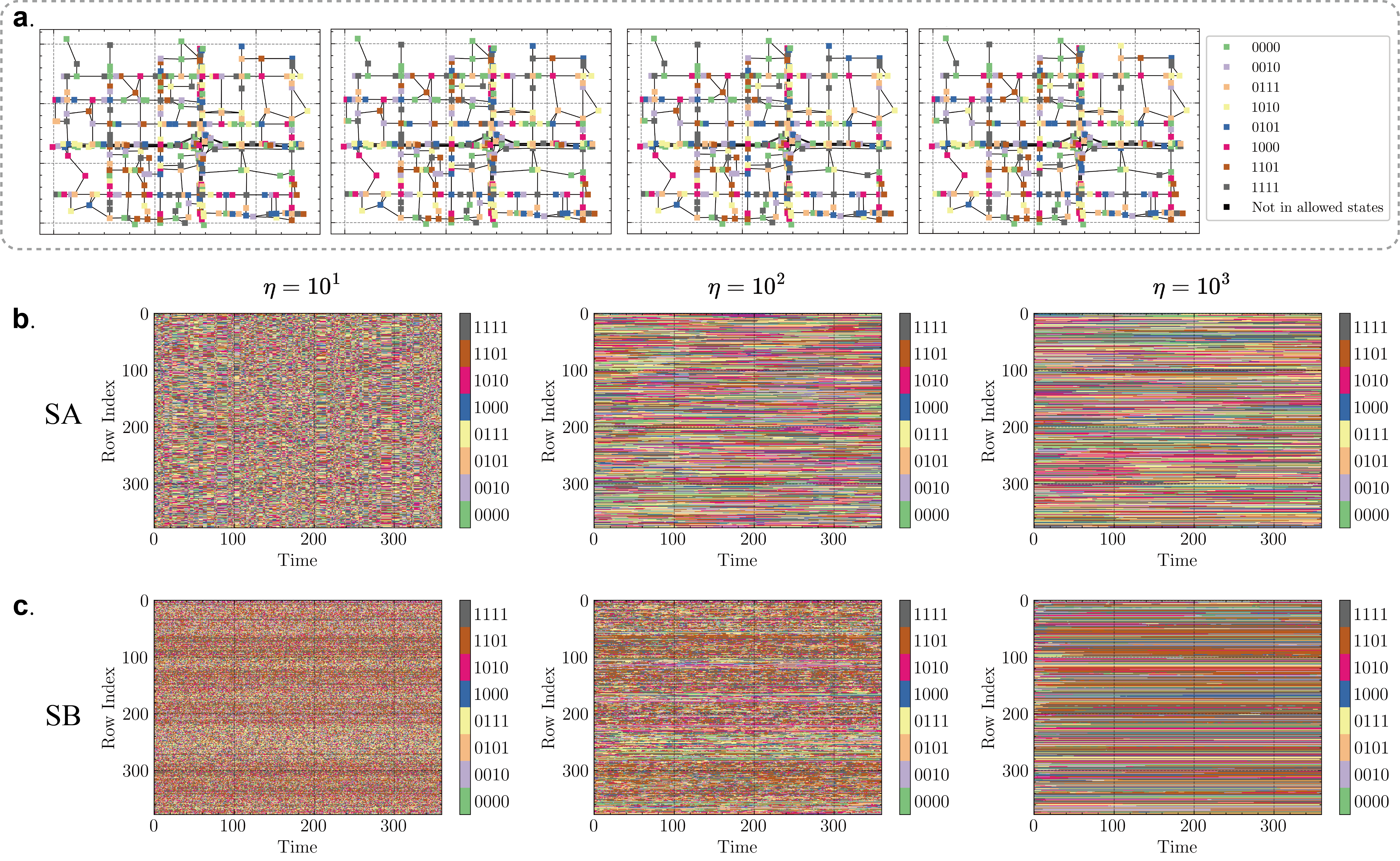}
        \caption{
            (a) Traffic light phase variation of different road nodes in the real road network at four consecutive time steps.
            (b) and (c) Show the state evolution sequences of the SA and SB algorithms at $\zeta = 10^{-4}$ for different values of $\eta$. Each row represents the time series of a road node, with $\eta = 10^1$, $\eta = 10^2$, and $\eta = 10^3$ from left to right.
        }
        \label{fig:Result-Timelist}
    \end{figure}

    In terms of solving efficiency, the SB algorithm outperforms the SA algorithm by effectively reducing the time complexity when solving with a computer and enhancing the model's ability to handle large-scale real-world road networks.
    Fig. \ref{fig:Result-Benchmark}(c) demonstrates a comparison of the solution efficiency of the two algorithms by randomly generating road topology networks of varying sizes and conducting tests using a single-core computation on an Intel i9-13900 CPU. 
    The results show that the time complexity of the SB algorithm is significantly lower than that of the SA algorithm.
    
    Specifically, the time complexity of the SA algorithm is approximately \(O(N^2)\), which exhibits typical characteristics of an NP-hard problem. 
    During the convergence process, the number of state transitions the SA algorithm needs to try is proportional to the number of nodes \(N\), and the time to calculate the Hamiltonian changes is also proportional to \(N\). 
    Therefore, as the network size \(N\) increases, the solving time for the SA algorithm grows quadratically with \(N\), leading to lower efficiency and even making it impractical for large-scale networks.    
    In contrast, the time complexity of the SB algorithm is approximately \(O(N^{1.35})\), primarily due to the optimization mechanism introduced by the momentum field \(\bm{y}\). 
    The momentum field effectively reduces the number of times the Hamiltonian with multi-body interactions needs to be recalculated during state changes, thus avoiding the high time complexity issue found in the SA algorithm. 
    By transforming the interactions between particles into interactions between particles and fields, the SB algorithm significantly reduces computational complexity, enabling more efficient solutions to NP-hard problems, especially in large-scale road networks.
    
    Fig. \ref{fig:Result-Timelist}(a) presents the evolution of the traffic signal states for four consecutive time steps during traffic flow optimization with the SB algorithm. 
    All traffic signals are able to accurately determine the corresponding phase states. 
    As time progresses, the traffic signal phases for most intersections remain consistent with the previous time step, while only a few intersections dynamically adjust their phases based on traffic flow changes to optimize the efficiency of the surrounding roads.    
    Fig. \ref{fig:Result-Timelist}(b) and Fig. \ref{fig:Result-Timelist}(c) further illustrate the time evolution for all nodes. 
    As \(\eta\) increases, it can be observed that the frequency of traffic signal state transitions for individual nodes gradually decreases, with the duration of the color bars in the heat map increasing. 
    A comparison of the results from the SA and SB algorithms reveals significant differences in their responses to changes in \(\eta\).

    \begin{figure*}[ht]
        \centering
        \includegraphics[width=\linewidth]{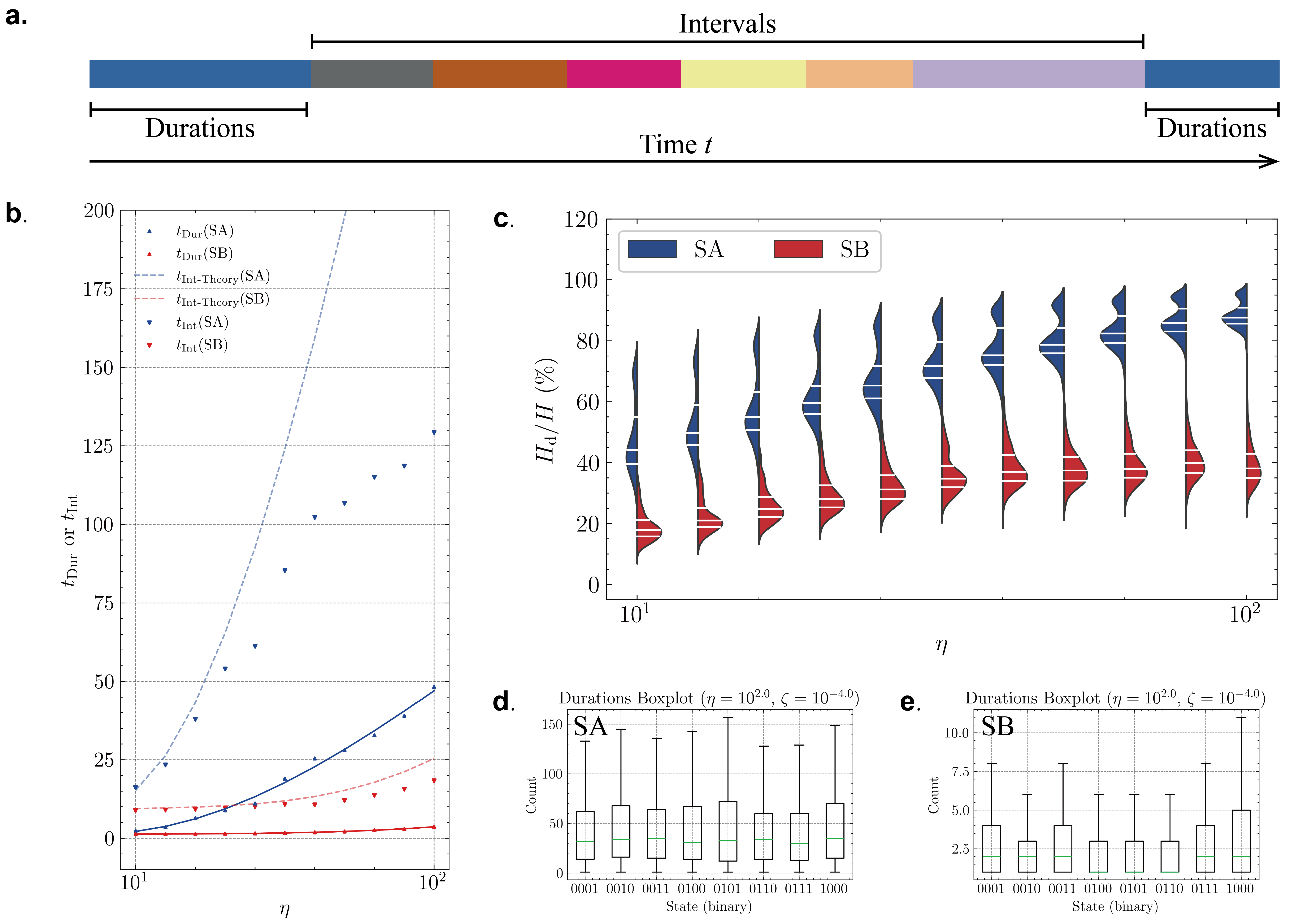}
        \caption{
            (a) Schematic illustrating how the Durations and Intervals of traffic light phase changes are computed.
            (b) Time-dependent variations of $t_{\text{Dur}}$ and $t_{\text{Int}}$ for the SA and SB algorithms, with units in iterations.
            (c) The proportion of $H_{\text{d}}$ in the total Hamiltonian under different values of $\eta$.
            (d, e) The distributions of durations for different traffic signal phases under the SA and SB algorithms, respectively.
        }
        \label{fig:Result-eta}
    \end{figure*}

    To further investigate the impact of parameter \(\eta\) on the frequency of phase switching at intersections, Fig. \ref{fig:Result-eta}(b) presents statistical analyses of the average duration time \(t_{\text{Dur}}\) and average interval time \(t_{\text{Int}}\) for each phase. 
    Here, \(t_{\text{Dur}}\) represents the duration of a single traffic signal phase, while \(t_{\text{Int}}\) refers to the waiting time from the end of one phase to the beginning of the next, as defined in Fig. \ref{fig:Result-Timelist}(a). 
    Ideally, the phases of each traffic signal should cycle periodically, meaning that the average interval time for each phase should be seven times its own duration, i.e., \(t_{\text{Int-Theory}} = 7t_{\text{Dur}}\).

    The results in Fig. \ref{fig:Result-eta}(b) show that, at smaller values of \(\eta\), the behavior of the SA and SB algorithms becomes similar. 
    This is because, at this point, the effect of \(H_{\text{d}}\) on the total Hamiltonian \(H\) is minimal, resulting in higher state switching frequency, which is reflected in \(t_{\text{Dur}} \to 1\). 
    This behavior indicates that, at low \(\eta\) values, the system tends to frequently switch phases in an attempt to find better solutions.
    As \(\eta\) increases, both the SA and SB algorithms show an upward trend in \(t_{\text{Dur}}\). 
    This is due to the increasing proportion of \(H_{\text{d}}\) in the total Hamiltonian, meaning that the cost of switching traffic signal phases becomes higher. 
    However, there is a significant difference in performance between the algorithms: \(t_{\text{Dur}}\) for the SA algorithm is generally higher than that of the SB algorithm, particularly at \(\eta = 10^2\), reflecting the difference in sensitivity to state switching during the optimization process between the two algorithms.    
    From Fig. \ref{fig:Result-eta}(c), which shows the statistical results for \(H_{\text{d}} / H_{\text{SB}}\), we see that the proportion of \(H_{\text{d}}\) in the SA algorithm is significantly higher than in the SB algorithm, reaching nearly 90\% at \(\eta = 10^{2.0}\). 
    At this point, the optimization of \(H_{\text{SB}}\) in the SA algorithm is almost entirely determined by \(H_{\text{d}}\), neglecting the regulation of road flow variance. 
    In contrast, for the SB algorithm, when \(\eta > 10^{1.5}\), the proportion of \(H_{\text{d}} / H_{\text{SB}}\) stabilizes around 30\%, showing a more stable optimization trend.    
    Furthermore, the relationship between \(t_{\text{Dur}}\) and \(\eta\) for the SA algorithm is nonlinear (as shown in Fig. \ref{fig:Result-eta}(b) with a logarithmic scale for \(\eta\)), indicating that the SA algorithm optimizes based solely on the specific value of \(H\) rather than using its derivative information. 
    Thus, the SA algorithm is more sensitive to the absolute values of the terms in \(H_{\text{SB}}\), making it difficult to smoothly adjust the system's convergence state via \(\eta\). 
    In contrast, the SB algorithm, by introducing the momentum field for optimization and constraining the system's initial state through potential wells, demonstrates a linear response to \(\eta\), enabling smoother regulation of traffic signal phases. 
    This phenomenon is consistent with the results in Fig. \ref{fig:Result-H}(b) and Fig. \ref{fig:Result-H}(c), confirming that the SB algorithm can more evenly optimize the weights of the terms in \(H_{\text{SB}}\), further verifying its smooth convergence characteristics when faced with varying \(\eta\) values.
    
    As \(\eta\) increases, the two algorithms begin to show distinct behaviors in \(t_{\text{Int}}\): for the SA algorithm at \(\eta = 10^{1.3}\), and for the SB algorithm at \(\eta = 10^{1.6}\), \(t_{\text{Int}}\) starts to deviate significantly from the theoretical value \(t_{\text{Int-Theory}}\). 
    This deviation can be understood as the algorithm's adaptive selection of high-frequency phases, which appear multiple times within a cycle, leading to an overall decrease in \(t_{\text{Int}}\). 
    For example, as shown in Fig. \ref{fig:Result-eta}(e), the occurrence frequency of phase 1000B is nearly twice that of phase 0100B.    
    However, from the optimization results of the SA algorithm (shown in Fig. \ref{fig:Result-eta}(d)), it is evident that the phases appear almost evenly distributed, failing to effectively optimize \(H_{\text{q}}\), which reflects the SA algorithm's shortcomings in optimizing high-frequency phase selections. 
    In contrast, the SB algorithm adapts more effectively to optimize the phase distribution, leading to better system performance.
    
    In summary, the value of \(\eta\) has a significant impact on both \(t_{\text{Dur}}\) and \(t_{\text{Int}}\), making it an effective way to adjust system behavior. 
    In real urban traffic systems, the flexible adjustment of this parameter is crucial. 
    Decision-makers can use this method to precisely control the frequency of traffic signal switches, optimize traffic flow, and reduce congestion. 
    Furthermore, \(\eta\) can not only be applied as a global parameter for the entire model, but could also be independently introduced for each traffic signal \(i\) as \(\eta_i\), enabling road-level control and urban traffic planning. 
    This idea has broad potential applications in urban traffic management, road-level control, and large-scale traffic system optimization.

%% file: Sections/Discussion.tex
The SB algorithm demonstrates a significant advantage over the SA algorithm in terms of convergence speed. 
The convergence curves in Fig. \ref{fig:Result-H}(a) show the behavioral differences between the two algorithms during different iterations, and these differences represent typical behaviors observed across multiple numerical experiments. 
From the convergence curve of the SA algorithm, we can observe that the system's Hamiltonian exhibits noticeable oscillations during the high-temperature phase. 
These oscillations reflect the random nature of the SA algorithm's search for microscopic states, as the system repeatedly attempts to explore multiple local optima. 
The search path lacks clear directionality, resulting in low search efficiency. 
This phenomenon is particularly prominent in the early stages of the algorithm, with oscillations gradually diminishing as the temperature decreases, eventually converging to a stable state. 
In contrast, the convergence process of the SB algorithm is much smoother. 
The \(H\) curve for the SB algorithm decreases progressively after approximately 20 iterations, showing a trend where the system's optimization state continually approaches the global optimum. 
This indicates that the SB algorithm is able to guide state evolution more efficiently during optimization, avoiding the chaotic search behavior seen in SA.
Overall, the optimization process of the SB algorithm is significantly superior to that of the SA algorithm. 
Its smooth convergence curve and flexible state transition mechanism enable it to handle complex system optimization problems more efficiently. 
This provides strong support for optimizing real traffic networks and reveals the potential advantages of the SB algorithm in combinatorial optimization.

Additionally, Fig. \ref{fig:Result-H}(b) and Fig. \ref{fig:Result-H}(c) present a comparison of the changes in \(H_{\text{q}}\) and \(H_{\text{d}}\) during the optimization process for both SA and SB algorithms. 
It is important to note that the \(H_{\text{q}}\) term is nonlocal in spatial dimensions, closely related to the interactions between neighboring intersections, but local in the time dimension, independent of specific time states. 
In contrast, the \(H_{\text{d}}\) term is local in space, depending only on the state of a single intersection, but has temporal correlations.
During the SA algorithm's optimization process, the system shows strong sensitivity to \(H_{\text{q}}\), with its proportion consistently above 60\%. 
This suggests that the SA algorithm prioritizes reducing traffic flow fluctuations, focusing mainly on optimizing flow variance while giving less attention to the continuity of states and their constraints. 
This tendency might lead to the neglect of dependencies between different states, causing the system to fail to fully meet long-term stability requirements in some cases. 
Moreover, the SA algorithm lacks sufficient consideration of the local temporal correlations, focusing too much on instantaneous local state changes and lacking a comprehensive understanding of the continuity of time evolution.

In contrast, during the optimization process of the SB algorithm, the weights of \(H_{\text{q}}\) and \(H_{\text{d}}\) are more balanced, with both terms contributing around 45\%. 
This indicates that the SB algorithm considers both traffic flow fluctuations and the continuity and constraints of states more evenly during optimization. 
The SB algorithm not only demonstrates strong capabilities in optimizing traffic flow variance but also better incorporates the local temporal state correlations, ensuring the system’s stability and consistency in global optimization.
Thus, the optimization process of the SB algorithm exhibits higher stability and greater sensitivity to the continuity parameter \(\eta\).

Furthermore, the SB algorithm also significantly outperforms the SA algorithm in terms of solving efficiency. 
Through tests on road networks of varying sizes, we find that the time complexity of the SB algorithm is much lower than that of the SA algorithm, especially for large-scale road networks.
For example, considering that there are approximately 9500 traffic signals in Beijing, the duration of a single red-green light phase typically ranges from \(10^1\) to \(10^2\) seconds. 
Ideally, a global optimization algorithm should be able to complete multiple optimizations within one signal phase cycle, allowing for real-time dynamic adjustments.
Based on the test results in Fig. \ref{fig:Result-Benchmark}(c), when using a single CPU core for optimization of the city's traffic signals, the computation time for the SA algorithm is estimated to be approximately \(2.05 \times 10^6\) seconds (about 24 days), while the SB algorithm requires only \(1.14 \times 10^4\) seconds (about 3 hours). 
Importantly, unlike the parallelization bottleneck of the SA algorithm, the SB algorithm, by introducing the momentum field mechanism, naturally possesses high parallel efficiency. 
Using multiple GPUs or FPGAs instead of a single CPU for acceleration, such as using 16 GPUs for parallel computation, the efficiency of the SA algorithm can be improved by about 100 times, while the SB algorithm's efficiency can be increased by a factor of 10,000 \cite{GotoHayato2021Highperformance}. 
With such accelerated conditions, the optimization time for the SA algorithm can be reduced to approximately \(10^4\) seconds (about 3 hours), while the optimization time for the SB algorithm can be further compressed to the level of seconds, reaching about 1 second. 
Furthermore, increasing the number of parallel GPUs or designing more efficient FPGA circuits can further enhance the system’s solution speed. 
This significant performance improvement makes the SB algorithm a practical solution for real-world traffic signal optimization systems.

In conclusion, the SB algorithm provides a powerful optimization tool for real-world traffic signal control systems, effectively handling complex optimization problems in large-scale traffic networks, particularly in real-time dynamic adjustments. 
Future research could explore even more efficient parallel computing strategies, such as FPGA-based hardware acceleration, to further improve its performance in large-scale traffic management applications.

%% file: Sections/Methods.tex
\subsection{Real Data Preparation}

    \begin{figure}[t]
        \centering
        \includegraphics[width=0.9\linewidth]{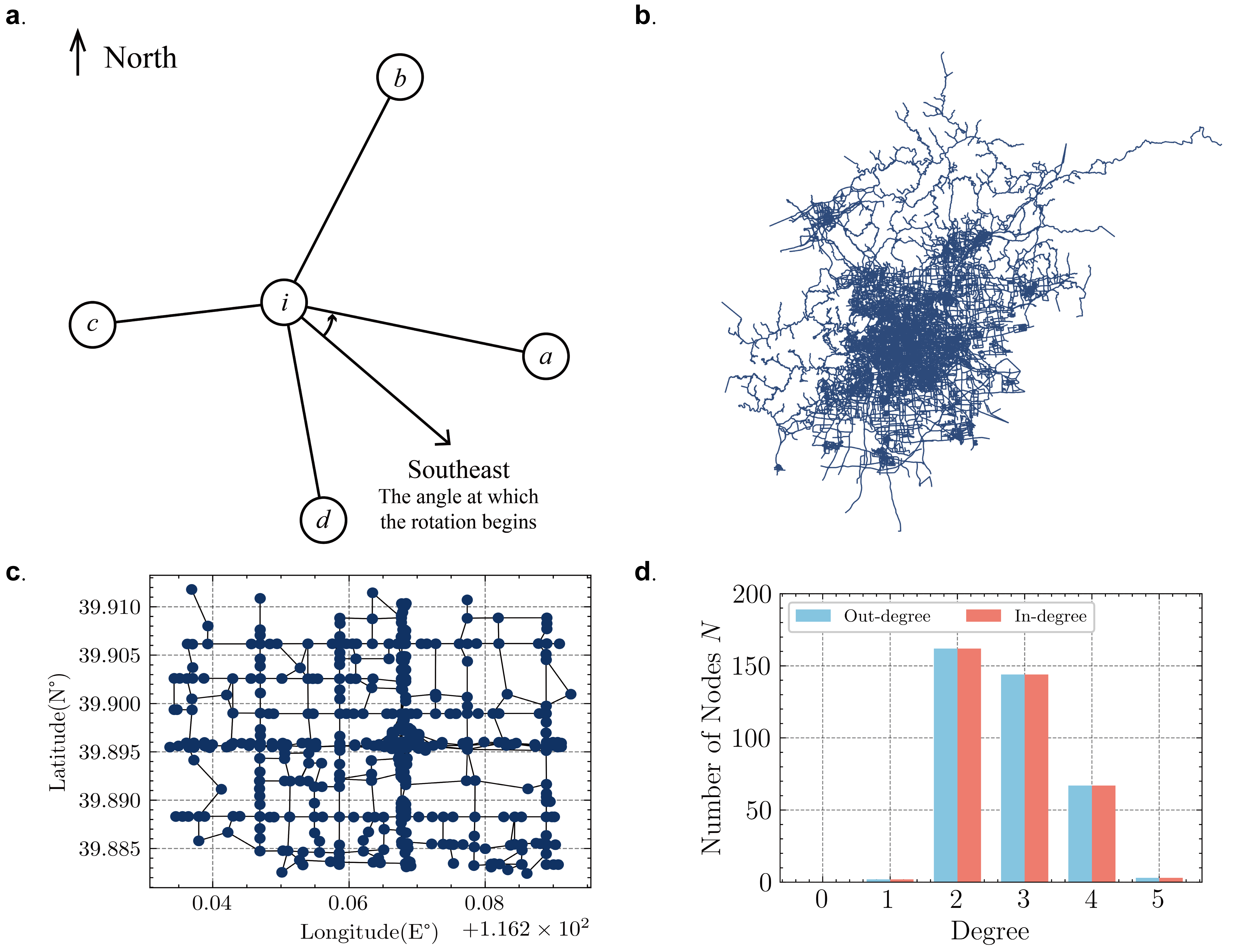}
        \caption{
            (a) Schematic of intersection quadrant retrieval.
            (b) Visualization of actual road network data from Beijing.
            (c) A subset of the real road network selected in this study.
            (d) Histogram of the in-degree and out-degree distributions of the road network.
            }
        \label{fig:RealRoad}
    \end{figure}

    To validate the feasibility of the proposed theoretical model and algorithm, this study constructs a spin-glass model of the actual road network based on real road data from Beijing for numerical experiments. 
    First, the topology of the traffic network is obtained: as shown in Fig. \ref{fig:RealRoad}(a), for any intersection \(i\), its southeastward \(45^\circ\) direction is defined as the base vector. Then, moving counterclockwise, the first encountered road direction is defined as the first quadrant road, corresponding to the east. Next, the roads in the second, third, and fourth quadrants are sequentially searched to define the direction of the roads.
    Considering that most urban road networks have fewer than four-way intersections, and these intersections are mostly overpasses and other types that do not require signal control, this study limits the number of intersections to four or fewer. In practical engineering applications, more intersections can be added according to the actual situation.
    
    As shown in Fig. \ref{fig:RealRoad}(b), the road network data used in this study comes from public data provided by Beijing's planning and construction department, combined with the 2023 OpenStreetMap (OSM) Beijing road network data. To control the computational load of the test network, we cropped the main road network of part of Beijing's urban area, covering the latitude and longitude range: longitude from \(116.23^\circ E\) to \(116.29^\circ E\) and latitude from \(39.88^\circ N\) to \(39.91^\circ N\). This dataset includes 393 traffic intersections and 647 roads. The cropped road network topology is shown in Fig. \ref{fig:RealRoad}(c). 
    We also counted the topological properties of the roads in the network and calculated the in-degree and out-degree of each intersection node in the traffic graph \(G\). Fig. \ref{fig:RealRoad}(d) shows the histograms of in-degrees and out-degrees of the nodes. The results indicate that, due to the well-connected structure of the network, the in-degree and out-degree of the nodes are almost identical (excluding one-way roads). Among them, nodes with in-degrees of 3 and 4 are the main targets of the optimization algorithm, while nodes with in-degrees of 1 and 2 serve as boundary nodes for periodic boundary treatment. Specifically, boundary nodes are randomly connected to form virtual intersections, thus creating connections that do not exist in the actual road network. The assumption of periodic boundary conditions is used only for numerical validation, ensuring the stability of the simulation. In real applications, such boundary conditions should be replaced by real-time information collected from larger-scale road sensors, such as speed sensors on urban boundary roads.
    
    In this study's experiments, we retrieved GPS signal data from vehicles traveling on actual roads between May 10, 2021, and May 11, 2021, via the Gaode API interface. By matching this data with road network data using longitude and latitude coordinates, traffic flow information is mapped onto the corresponding roads, resulting in real traffic flow change sequences \(\{q_{i,j}(t) | (i,j) \in E\}\) and turning probabilities for each intersection.
    The specific estimation process is as follows:
    To compute the turning probabilities associated with road \((j,i)\), assume its left turn connects to road \((i,l)\), and the right turn connects to road \((i,r)\). The traffic flow change on road \((i,j)\) can be computed from the actual data:
    \begin{equation}
        \Delta q_{i,j} = [q_{i,j}(t) - q_{i,j}(t-1)]/\varsigma_{i,j},
    \end{equation}
    where \( q_{i,j}(t) \) and \( q_{i,j}(t-1) \) represent the traffic flow on road \((i,j)\) at times \( t \) and \( t-1 \), respectively, and \( \varsigma_{i,j} \) represents the number of lanes on road \((i,j)\). Lane normalization of \( \varsigma_{i,j} \) ensures the comparability of traffic flow across different roads.
    Left and right turn probabilities can be estimated based on the flow changes:
    \begin{equation}
        \alpha_{j,i} = \Delta q_{i,l}(t-1) / \Delta q_{j,i}(t-1),
    \end{equation}
    \begin{equation}
        \beta_{j,i} = \Delta q_{i,r}(t-1) / \Delta q_{j,i}(t-1).
    \end{equation}
    
    However, the above estimates are clearly subject to some error. This case is mainly used to demonstrate the effectiveness of the optimization algorithm. The model established for the real road network and road conditions is intended to show the advantages of the optimization algorithm in large-scale intersections and multi-phase signal control, capable of dynamically responding to traffic pattern changes, including peak traffic fluctuations, adverse weather impacts, and differences between holiday and workday travel, achieving global optimization in complex scenarios.
    It should be noted that, for practical deployment of this model, traffic flow and road turning data can be predicted through large-scale neural network models \cite{Guo2019Attention, S.Guo2019Deep, Tang2019Traffic, Wu2018hybrid, Xiao2019Hybrid}, or real-time estimation can be done by deploying sensor devices such as image sensors and speed sensors, which are already installed in many large cities, to obtain more accurate real-time traffic conditions.

\subsection{Numerical Experiment Parameters}
    To ensure the comparability of the solution processes between different algorithms, we adopted a unified objective function \( H \) as the evaluation criterion, as shown in equation (\ref{eq:H-base-fix}) in the tests.
    Where the \( H_{\text{d}} \) term is related to the state from the previous time step, \( \bm{x}(t-1) \). Specifically, at the initial time step \( t = 0 \), \( \bm{x}(t-1) \) is randomly selected from the set \( M \). For any subsequent time step \( t > 0 \), \( \bm{x}(t-1) \) refers to the state that minimizes the Hamiltonian \( H(\bm{x}(t)) \) at the previous time step, i.e.,
    \begin{equation}
        \bm{x}(t-1) = \arg \min_{\bm{x} \in \mathcal{X}} H[\bm{x}( t-1),\bm{x}( t-2), t-1]
    \end{equation}
    where \( \mathcal{X} \) represents the entire possible state space, and \( H(\bm{x}, t-1) \) is the Hamiltonian of the current state \( \bm{x} \) at time \( t-1 \).    
    In the computational tests, the parameters for the SA algorithm were set as \( \omega = 0.2 \), the initial temperature \( T_{\text{Init}} = 10^1 \), and the final temperature \( T_{\text{End}} = 10^{-4} \). 
    For the SB algorithm, the parameters were set with the initial potential well strength \( a_0 = 1 \), a maximum number of iterations \( Iter_{\text{max}} = 2000 \), and a precision control parameter \( \zeta = 10^{-4} \).     
    These parameter choices ensure that both algorithms run under the same testing conditions, allowing for a fair comparison of their performance when applied to real traffic data.

%% file: SupplementarySections/SS1.tex
The global optimization of traffic signals is widely recognized as a combinatorial optimization problem within the academic community \cite{Eom2020traffic, Shaikh2022Review, Tchuitcheu2020Internet}. However, from both mathematical and artificial intelligence algorithmic perspectives, it remains one of the most challenging problems.
By exploring its potential connections with physical models such as the Ising model, these problems can be reformulated as statistical physics problems. This approach leverages methodologies from physics to provide a novel pathway for solving combinatorial optimization problems.
In this context, Suzuki et al. were the first to propose a research framework combining traffic signal optimization with the Ising model, revealing its potential applications in physical modeling \cite{Suzuki2013Chaotic}. 
This section focuses on the methodology for mapping the global optimization of traffic signals onto the universal Ising model framework, laying a theoretical foundation for future extensions to complex road network structures and multi-phase traffic signal scenarios.

\subsection{Combinatorial Optimization for Traffic Signals}

    Combinatorial optimization problems are a class of optimization tasks centered on discrete variables, aiming to identify the optimal combination of variables under a set of constraints. Mathematically, a combinatorial optimization problem can be defined as \cite{Cook1994Combinatorial}:
    \begin{equation}
        \min_{\bm{x} \in S} f(\bm{x}),
        \label{eq:min_f}
    \end{equation}
    where \(\bm{x} = [x_1, x_2, \dots, x_n]\) represents a decision vector defined on a discrete set \(D\); \(S \subseteq D^n\) denotes the feasible solution space constrained by specific conditions; and \(f(\bm{x})\) is the objective function used to evaluate the quality of variable combinations.

    In general, any complex nonlinear function can be expressed as a summation of terms with different orders through techniques such as Taylor expansion or Fourier transform \cite{Boik2008implicit, Bulos2005modified, Mouroutsos1985Taylor, Nourazar2015new}. This decomposition is not only universal but also provides deep insights into the behavior of the function. Therefore, for any objective function \(f(\bm{x})\) involving the decision vector \(\bm{x}\), it can be represented as:
    \begin{equation}
        f(\bm{x}) = \sum_{i} h_i x_i + \sum_{i,j} J_{ij} x_i x_j + \sum_{n=3}^{\infty} \sum_{i_1, i_2, \dots, i_n} K_{i_1 i_2 \dots i_n} x_{i_1} x_{i_2} \dots x_{i_n},
        \label{eq:CO}
    \end{equation}
    where the first term \(\sum_{i} h_i x_i\) represents the linear contribution of individual variables, with \(h_i\) reflecting the direct influence of each variable on the objective function. The second term \(\sum_{i<j} J_{ij} x_i x_j\) accounts for pairwise interactions between variables, where \(J_{ij}\) quantifies the strength of the interaction between \(x_i\) and \(x_j\). These two terms often dominate the system's behavior. Higher-order interaction terms describe the complex relationships among three (\(n=3\)) or more (\(n>3\)) variables, with \(K_{i_1 i_2 \dots i_n}\) capturing the strength and impact of these interactions. For systems with multi-variable coupling, these terms are essential. This representation is widely applicable, especially in optimization and modeling, where truncating higher-order terms simplifies computations while preserving critical system behavior.

    In the global optimization of traffic signals, the state of each intersection's traffic signals can be represented as discrete variables, such as \(x_i \in \{-1, +1\}\), corresponding to two signal phases: east-west and north-south traffic flows. The objective is to optimize the combination of signal states to maximize traffic flow and minimize average waiting times. Thus, the objective function for traffic signal optimization can be formalized similarly to Equation (\ref{eq:CO}):
    \begin{equation}
        f(\bm{x}) = \sum_{i} h_i x_i 
        + \sum_{i} J_{ii} x_i^2
        + \sum_{i \neq  j} J_{ij} x_i x_j 
        + O\left(\sum_{i_1, i_2, \dots, i_n} K_{i_1 i_2 \dots i_n} x_{i_1} x_{i_2} \dots x_{i_k}, n \geq 3 \right),
        \label{eq:Ising}
    \end{equation}
    where the linear term \(\sum_{i} h_i x_i\) represents the bias for individual signals, with \(h_i\) reflecting the priority or external influence at intersection \(i\), such as traffic demand or regional importance. This term primarily impacts local traffic signal decisions.
    The second-order self-interaction term \(\sum_{i} J_{ii} x_i^2\) describes temporal self-correlation effects of traffic signals at an intersection, accounting for how current traffic flow affects future flow at the same location.
    The pairwise interaction term \(\sum_{i \neq j} J_{ij} x_i x_j\) captures the coordination requirements between traffic signals at different intersections. \(J_{ij}\) quantifies the weight of synchronization, such as green wave coordination or flow consistency between neighboring intersections. This term is critical for optimizing overall traffic flow through signal interdependence.
    The higher-order interaction term \(O\left(\sum_{i_1, i_2, \dots, i_n} K_{i_1 i_2 \dots i_n} x_{i_1} x_{i_2} \dots x_{i_k}, n \geq 3 \right)\) represents complex interactions among three or more signals, such as strong local coupling at certain intersections or global traffic coordination effects. This term is particularly useful for modeling multi-intersection dependencies and capturing complex behaviors in large-scale traffic networks.
    Together, these terms allow the cost function to accurately describe the various relationships and requirements in traffic signal optimization, providing a robust mathematical foundation for global optimization.

\subsection{Ising Model}

    The Ising model, one of the simplest and most fundamental models in the field of statistical mechanics, was originally developed to describe the interactions between discrete variables in lattice spin systems \cite{Brush1967History, Cipra1987introduction}. Over time, its applications have extended far beyond statistical mechanics, making a profound impact in fields such as theoretical physics, artificial intelligence \cite{Aoki2016Restricted, Gu2022Thermodynamics, Yoshioka2019Transforming} and combinatorial optimization \cite{Bybee2023Efficient, Dan2020Clustering, Zhang2022review}.
    In particular, the Ising model directly inspired the development of the Boltzmann machine, providing the foundational energy function and binary state representation that enabled the network to learn and identify complex data features, marking a crucial component of the advancements in machine learning that led to the 2024 Nobel Prize in Physics.

    \begin{figure}[tp]
        \centering
        \includegraphics[width=\textwidth]{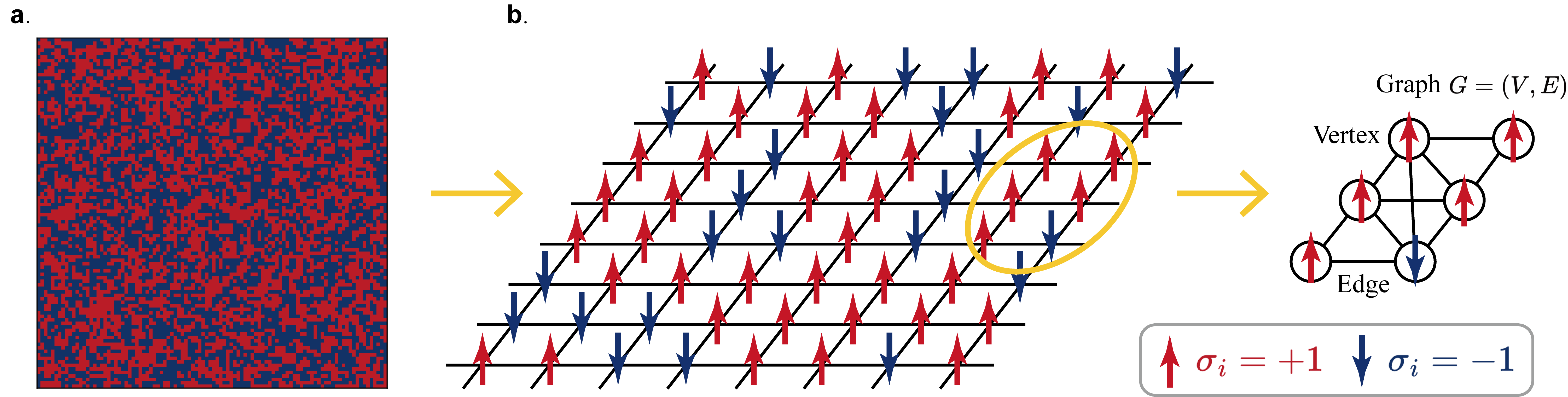}
        \caption{
            a. A result diagram of a steady-state two-dimensional Ising model with \(10^4\) spins.  
            b. Schematic diagram of the Ising model.  
        }
        \label{fig:IsingModel}
    \end{figure}

    The core of the Ising model lies in its ability to represent the macroscopic behavior of a system composed of an enormous number of microelements (spins), which interact with each other and respond to external forces. These interactions are mathematically captured through a defined coupling relationship between spins and external biases. This framework leads to the construction of an energy function, commonly referred to as the Hamiltonian, which characterizes the system's state. By minimizing this energy function, the model effectively identifies the system's equilibrium state or optimizes the target function.
    
    The Ising model is typically defined on an undirected graph \(G = (V, E)\), where \(V\) is the set of vertices and \(E\) is the set of edges. Each microelement is called a 'spin'. A spin variable exists on each vertex and is expressed as \(\sigma_i\), where \(\sigma_i = \pm 1\).
    
    It is necessary to express the cost function of a combinatorial optimization problem and its constraints using the Hamiltonian (energy function) of the Ising model:
    \begin{equation}
        H_{\text{Ising}}(\{\sigma_i\}) = \sum_{i \in V} h_i \sigma_i + \sum_{(i, j) \in E} J_{ij} \sigma_i \sigma_j,
    \end{equation}    
    where \(h_i\) represents the magnetic field applied to the site \(i \in V\), \(J_{ij}\) denotes the interaction between sites \(i\) and \(j\) for \((i, j) \in E\).
    In terms of physics, the minimum energy state is referred to as a stable state.
    
\subsection{Solving Combinatorial Optimization Problems with the Ising Model}

    By comparing the formal representation of combinatorial optimization problems (such as Equation (\ref{eq:CO})) with the mathematical expression of the Ising model (Equation (\ref{eq:Ising})), a significant structural similarity can be observed between the two. Both can be reduced to the summation of linear and quadratic terms over a set of discrete variables. This correspondence in mathematical form allows the Ising model to effectively capture the interdependencies between variables in discrete systems and the global optimization objective.    
    This modeling approach offers new perspectives for solving complex combinatorial optimization problems, such as the maximum cut problem, graph coloring problem, and path planning problem \cite{Dan2020Clustering}.
    
    The advantages of the Ising model are not only reflected in its flexibility in problem representation but also in its efficiency in practical applications. Traditional optimization algorithms (such as simulated annealing and genetic algorithms) can be used to solve the energy minimization problem of the Ising model. However, using the inherent dynamics of physical systems for natural evolutionary solutions offers additional advantages to this model.    
    For instance, quantum annealing technology leverages the evolution mechanisms of quantum systems to rapidly find the energy-minimal solution \cite{Inagaki2016coherent}, while optical computing hardware enables large-scale parallel computation of variables, significantly improving computational efficiency \cite{Yamaoka201520kspin}.     
    With advances in algorithms, heterogeneous computing technologies such as optical chips, and improvements in physical system simulation capabilities, the practical application of the Ising model will have even broader prospects, providing important insights for scientific research and engineering practices.

\subsection{Example: Grid-Based Road Network and the Ising Model}

    \begin{figure}[t]
        \centering
        \includegraphics[width=0.7\textwidth]{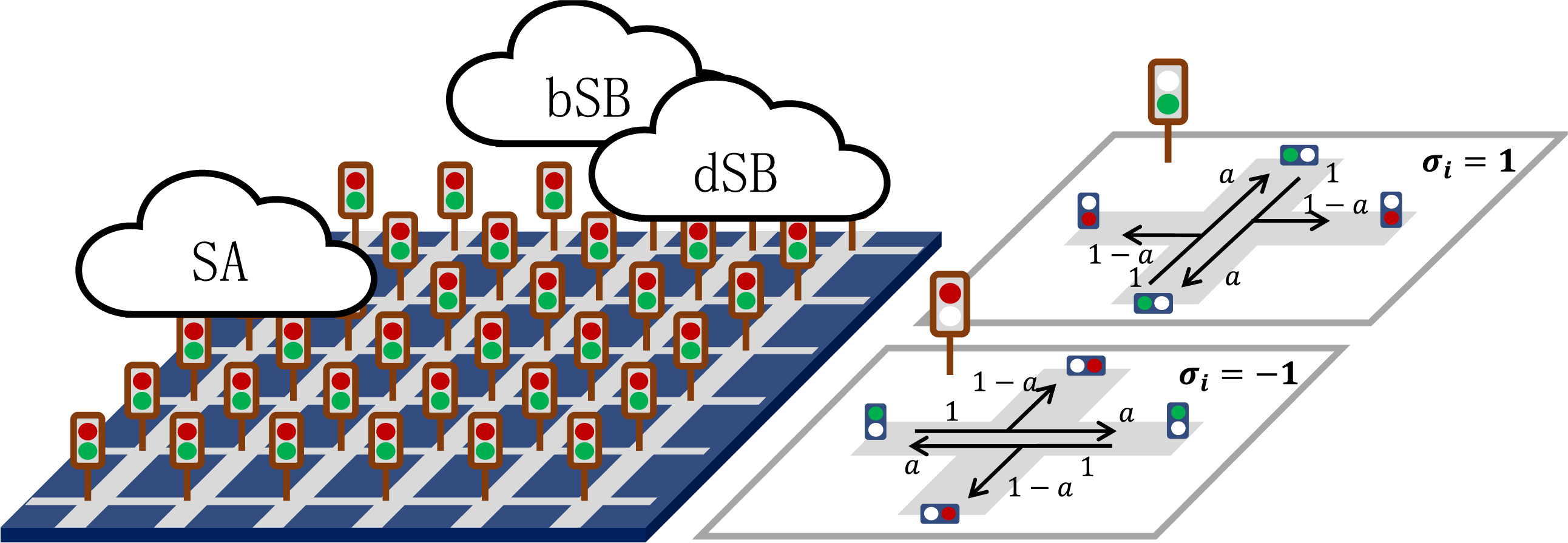}
        \caption{Schematic diagram of traffic flow in a grid-based road network.}
        \label{fig:trac}
    \end{figure}

    In the traffic signal optimization problem, the flow variance serves as the cost function, reflecting the imbalance in traffic flow across the entire road network. A higher variance indicates that traffic flow in some directions at intersections is either too high or too low, reducing overall traffic efficiency. Therefore, calculating the flow variance of the traffic system provides an effective measure of the system's efficiency. Some works have focused on the similarity between the flow variance function used in traffic signal optimization and the Hamiltonian of the Ising model, providing rigorous mathematical proofs \cite{DaisukeInoue2021Traffic, Suzuki2013Chaotic}.

    Specifically, as shown in Fig. \ref{fig:trac}, consider an \(L \times L\) (where \(L \in \mathbb{N}\)) road network, where east-west and north-south roads intersect with periodic boundary conditions. Each road consists of two lanes, one for each direction. At each intersection, a traffic signal controls the flow of vehicles. The signal state \(\sigma_i\) at each intersection can have two possible states: \(\sigma_i = +1\), allowing traffic to flow only in the north-south direction; and \(\sigma_i = -1\), allowing traffic to flow only in the east-west direction. Each vehicle passes through the intersection with a fixed probability \(a \in [0, 1]\). If the vehicle does not go straight, it turns left or right with equal probability, i.e., the turning probability is \((1-a)/2\).

    The evolution equation for the number of vehicles \(q_{ij} \in \mathbb{R}^+\) from intersection \(j\) to intersection \(i\) is given by the following difference equation:
    \begin{equation}
        q_{ij}(t + 1) = q_{ij}(t) + s_{ij} \cdot \frac{1}{2} (-\sigma_i + \alpha \sigma_j),
        \label{eq:q_ij}
    \end{equation}
    where \(\alpha \equiv 2a - 1\), and \(s_{ij} \in \{ \pm 1 \}\) represents the lane direction from intersection \(j\) to intersection \(i\); \(s_{ij} = +1\) represents the north-south direction, and \(s_{ij} = -1\) represents the east-west direction. Note that \(q_{ij}\) is normalized to represent the number of vehicles passing through the lane per unit time.

    Next, the deviation between the north-south and east-west traffic flows at each intersection \(i\) can be defined as:
    \begin{equation}
        \kappa_i(t) \equiv \sum_{j \in N(i)} s_{ij} q_{ij}(t) \cdot \frac{1}{2},
    \end{equation}    
    where \(N(i)\) represents the indices of the four neighboring intersections of intersection \(i\). Equation (\ref{eq:q_ij}) converts this into the time evolution equation for the flow deviation \(\kappa(t)\):
    \begin{equation}
        \bm{\kappa}(t+1) = \bm{\kappa}(t) + 4 (-\bm{I} + \alpha)  \bm{A} \bm{\sigma}(t),
    \end{equation}
    where the flow deviation vector is defined as \(\bm{\kappa } \equiv [\kappa_1, \kappa_2, \dots, \kappa_{L^2}]^\top\), the signal state vector is defined as \(\bm{\sigma} \equiv [\sigma_1, \sigma_2, \dots, \sigma_{L^2}]^\top\), and \(\bm{A} \in \mathbb{R}^{L^2 \times L^2}\) is the adjacency matrix of the periodic grid graph, with \(\bm{I}\) being the identity matrix.

    This allows the definition of an objective function to assess the traffic conditions at each time step:
    \begin{equation}
        H_{\text{Ising-Traffic}}[\bm{\sigma}(t)] \equiv \bm{\kappa }(t+1)^\top \bm{\kappa }(t+1) + \eta [\bm{\sigma}(t) - \bm{\sigma}(t-1)]^\top [\bm{\sigma}(t) - \bm{\sigma}(t-1)].
        \label{eq:H_ising}
    \end{equation} 
    The first term on the right-hand side suppresses the flow deviation at each intersection for the next time step, while the second term restricts the frequent switching of signal states at each intersection. The parameter \(\eta \in \mathbb{R}^+\) is used to determine the weight ratio between the two terms. 

    According to the general form of the combinatorial optimization problem given by Equation (\ref{eq:min_f}), the objective function (\ref{eq:H_ising}) needs to be minimized, i.e., finding \(\bar{\sigma}(t)\) that minimizes this objective function:
    \begin{equation}
        \overline{\boldsymbol{\sigma}}(t)=\underset{\boldsymbol{\sigma} \in\{ \pm 1\}^{L^2}}{\operatorname{argmin}} H[\boldsymbol{\sigma}(t)] .
    \end{equation}

    It is noteworthy that the objective function can be rewritten in the form of the Ising model as \cite{DaisukeInoue2021Traffic}:
    \begin{equation}
        H_{\text{Ising-Traffic}}[\boldsymbol{\sigma}(t)] = \bm{\sigma}(t)^{\top} \bm{J} \bm{\sigma}(t) + \bm{h} \bm{\sigma}(t),
    \end{equation}
    where \(\bm{J}\) represents the coupling matrix between nodes, and \(\bm{h}\) is the external field bias vector, specifically:
    \begin{gather}
        \bm{J}\equiv\Big(-\bm{I}+\frac{\alpha}{4}A\Big)^{\top}\Big(-I+\frac{\alpha}{4} \bm{A} \Big)+\eta \bm{I} ,
        \\
        \bm{h}\equiv2\bm{\kappa}(t)^T\Big(-\bm{I}+\frac{\alpha}{4}\bm{A}\Big)-2\eta\bm{\sigma}(t-1)^T  .
    \end{gather}
    Thus, the traffic signal optimization problem is equivalent to solving the Ising problem for the lowest energy state in a two-dimensional spin system.

    \subsection{Spin-Glass Model}

    The spin-glass model is a class of spin systems with glass-like properties, which not only has significant value in the study of amorphous materials and glass transitions but is also widely applied in combinatorial optimization problems, constraint satisfaction problems, and other complex system studies in computer science \cite{Katzgraber2008New, Zhu2016Brief}. For example, it has been used in error correction coding and decoding problems in information science \cite{Chowdhury2014Spin}, image restoration and compressed sensing problems \cite{Inoue2003Image, Lustig2007Sparse}, associative memory and perceptron learning problems in artificial intelligence \cite{Rosen1993Large, Goodwin1988Exploration}, network structure prediction and reconstruction, network community structure partitioning problems in complex network science \cite{Metz2021Mean-field}, and protein structure prediction problems in life sciences \cite{Friedrichs1989Toward, Goldstein1992Optimal}. These applications can all be transformed into spin-glass systems for study.
    Research in this field has led to fruitful outcomes, promoting interdisciplinary integration and fusion across fields such as statistical physics, computer science, information theory, and complex systems research.

    \begin{figure}[tp]
        \centering
        \includegraphics[width=0.5\textwidth]{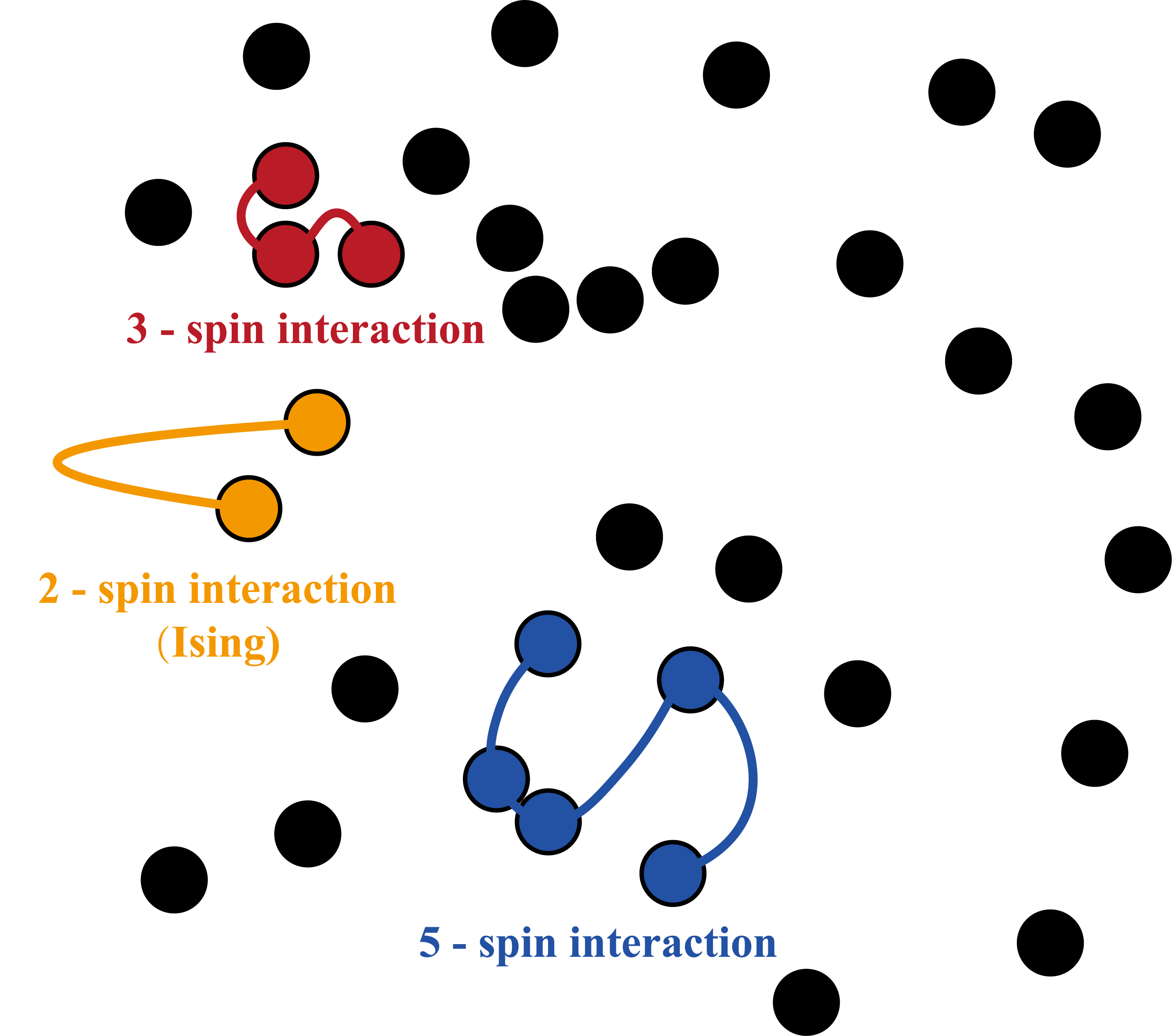}
        \caption{Schematic diagram of the spin-glass model.}
        \label{fig:SpinGlassModel}
    \end{figure}

    Depending on the relationships between different nodes, the spin-glass model can take various forms. As shown in Fig. \ref{fig:SpinGlassModel}, the Hamiltonian \(H_{\text{p-Spin}}\) of a \(p\)-spin glass model, which involves interactions between \(p\) spins, is given by \cite{kirkpatrick1987p}:
    \begin{equation}
        H_{\text{p-Spin}} = -\sum_{1\leqslant i_1 < i_2 < ... < i_p \leqslant N} J_{i_1 i_2 ... i_p} x_{i_1} x_{i_2} ... x_{i_p} + \sum_{i=1}^{N} h_i x_i,
        \label{eq:SpinGlass}
    \end{equation}
    Here, the model contains \(N! / [p!(N-p)!]\) coupling constants \(J_{i_1 i_2 ... i_p}\), which are independent and represent the interaction strength between \(p\) different nodes \(x_{i_1} x_{i_2} ... x_{i_p}\). The coefficient \(h_i\) represents the external field acting on node \(x_i\). Overall, the total potential energy from the many-body interactions between all nodes and the total potential energy from the external fields acting on each node compose the system's Hamiltonian. As in the Ising model, \(x_i = \pm 1\).

    Compared to Equation (\ref{eq:CO}), Equation (\ref{eq:SpinGlass}) reveals the main distinction between the spin-glass model and the Ising model: it introduces many-body interactions between spins, making the system more complex and disordered. The spin-glass model is a more generalized mathematical form of combinatorial optimization problems, with the core goal being to optimize the system's Hamiltonian to find the global optimal solution that satisfies the constraints.
    In practical urban traffic network optimization, the spin-glass model can more effectively represent the cooperative control of multi-intersection signal systems, accurately capturing the interdependencies between complex intersections. This provides strong modeling support for optimizing multi-phase signal systems, showing significant advantages in improving the precision of traffic signal optimization and adapting to complex road network environments.

%% file: SupplementarySections/SS2.tex
Based on the traditional definition of road intersections in signal control systems, this study uses the widely adopted four-way intersection with eight signal phases as the basic intersection model to represent urban road networks. Additionally, to reflect the topological structure of real-world road connections, the study employs a graph data structure, which not only aids in abstracting and simplifying the road network but also lays the foundation for subsequent analysis and optimization.
        
The traffic signals of an eight-phase intersection can be seen as a combination of basic steering patterns from each road direction. Since most intersections allow simultaneous right turns and straight movements, each direction's left turn, as well as the simultaneous straight and right turns, are defined as a basic steering direction (Basic steering direction encoding). Based on this classification, as shown in Fig. 1(a) in the main text, eight basic turning patterns are derived as the fundamental units for modeling.

To describe the correlation between different phases of an eight-phase intersection (for example, the full green phase and the straight-right-turn phase, which both allow straight and right turns), we use a two-bit binary number combined with a two-bit mask (XX) to encode the basic steering direction. This encoding method has the advantage of directly associating traffic flow variations with specific binary bits, providing a clear mathematical framework for the subsequent flow evolution equations. This method efficiently represents traffic flow characteristics with similar turning phases, allowing for a more effective description and optimization of the traffic signal switching process. Additionally, this encoding reduces the potential barrier between phases with similar traffic flow patterns; for example, transitioning from \(1101_B\) to \(1111_B\) only requires overcoming a single bit barrier. In the optimal state search, this characteristic enables smooth gradient transitions, improving the convergence and stability of numerical algorithms and providing more precise parameter support for traffic optimization.

A typical intersection has four directions, and we can combine the basic turning directions without traffic conflicts to form pairwise combinations, establishing a model to describe the well-known four-way intersection with eight signal phases. To map the real-world network model to the general form of the spin-glass model, we further assume that the state of each traffic signal is controlled by a four-bit binary number (as shown in Fig. 1(b) in the main text). Specifically, the state of the signal is defined as a four-bit binary number, expressed as \(x^{(4)}x^{(3)}x^{(2)}x^{(1)}\mathrm{B}\), where each \(x^{(i)}\) (\(i = 1, 2, 3, 4\)) takes a value of 0 or 1, representing the binary value at that position, and the suffix B indicates binary counting. This four-bit representation efficiently expresses the changes in signal states and corresponds to the state variables in the spin-glass model, making the simulation of traffic signal states more concise and practical.

In addition to the four-way intersection, T-junctions also account for a significant proportion of real-world urban road networks. Therefore, it is necessary to consider the modeling of T-junctions and ensure that the four-way intersection with eight signal phases model is compatible with the traffic rules of T-junctions. The primary difference between a T-junction and a four-way intersection is that a T-junction lacks one direction, with only three traffic signal directions.        
Therefore, to ensure that traffic flow at T-junctions and four-way intersections is accurately represented within a unified traffic signal optimization framework, the signal phases of T-junctions need to be mapped to the phases of the four-way intersection with eight signal phases model. The primary function of this mapping is to establish the correspondence between the eight phases of a four-way intersection and the various phases of a T-junction, ensuring consistency between the two types of intersections in traffic signal control.

In this study, the most common three-phase pattern is used to model the operational mode of T-junctions. To accommodate the traffic rules of T-junctions, we map the eight-phase model of the four-way intersection to the three-phase model of the T-junction. Specifically, Fig. 1(c) in the main text shows the encoding method for the basic steering rules of a T-junction, with unused turning encodings from the four-way intersection discarded (marked with red crosses). Fig. 1(d) in the main text demonstrates the three valid phases of operation for a T-junction.        
It is important to note that, due to the lack of one direction in a T-junction, its three signal directions can actually correspond to any of the four intersection quadrants. Therefore, the three-phase model for a T-junction has 12 possible configurations, which can be obtained by rotating the three-phase road layout shown in Fig. 1(d) in the main text clockwise by 90, 180, and 270 degrees. To achieve this, we use basic steering direction encoding to map the three-phase models of T-junctions with different orientations, ensuring that each scenario is accurately represented. The specific phase mapping rules can be found in Table (\ref{tab:01}), which lists the encoding correspondence between the three-phase model of various T-junction orientations and the four-way intersection with eight signal phases model.        
Through this approach, we ensure the validity of T-junctions in traffic signal control, while also providing clear and accurate parameter support for optimizing traffic flow between different types of intersections.

\begin{sidewaystable}
    \caption{Phase Correspondence Table for Cross and T-Junction Intersections}
    \label{tab:01}
    \begin{tabular*}{\textwidth}{@{\extracolsep\fill}lccccccccccccccc}
        \toprule
        & \multicolumn{3}{@{}c@{}}{East} & \multicolumn{3}{@{}c@{}}{North} & \multicolumn{3}{@{}c@{}}{West} & \multicolumn{3}{@{}c@{}}{South} \\
        \cmidrule{2-4} \cmidrule{5-7} \cmidrule{8-10} \cmidrule{11-13}
        Phase of T-junction & I & II & III & I & II & III & I & II & III & I & II & III \\
        \midrule
        Related Intersection Phase & III,VII,VIII & IV,V & I,VI &
        II,IV,VI & I,VIII & V,VII &
        III,VII,VIII & IV,V & I,II &
        II,IV,VI & I,VIII & III,V \\
        \bottomrule
    \end{tabular*}
\end{sidewaystable}

%% file: SupplementarySections/SS3.tex
\begin{figure}[tp] 
    \centering
    \includegraphics[width=\linewidth]{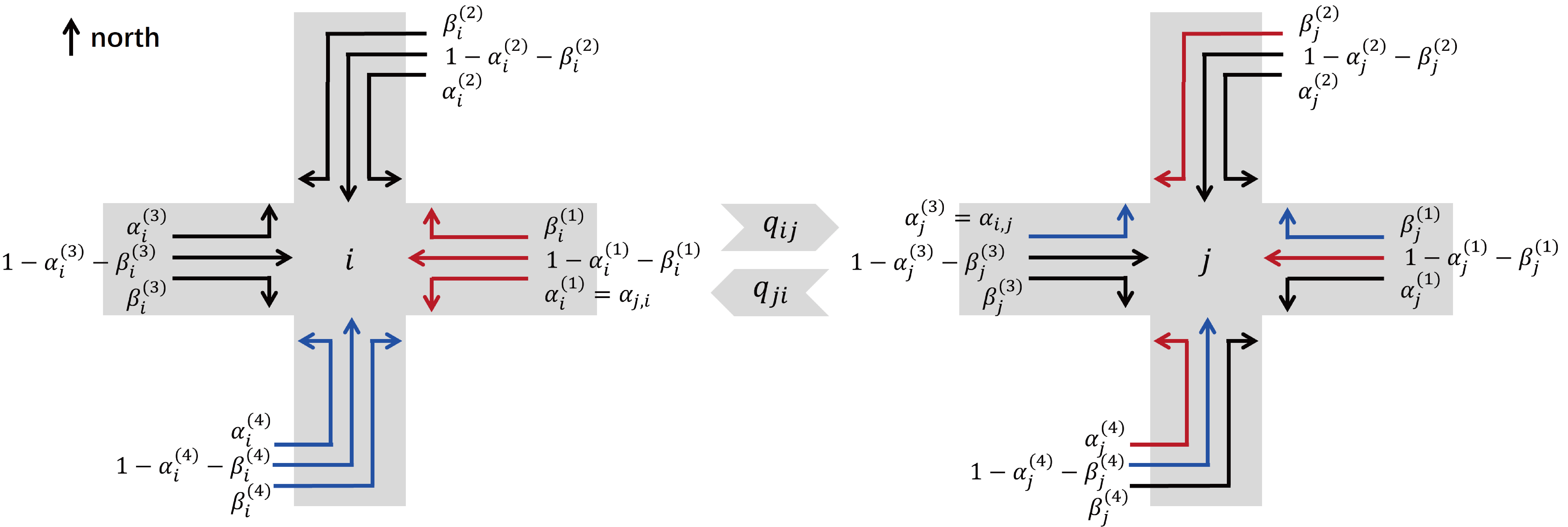}
    \caption{
        Schematic diagram of a four-way, eight-phase intersection in a real road network model, where red and blue arrows represent the vehicle flow conditions when road $(i,j)$ is connected in the east-west and north-south directions, respectively (traffic flows from intersection $j$ into intersection $i$).
        }
    \label{fig:TrafficFlow}
\end{figure}

Assume that the road network we are dealing with consists of $N$ traffic intersections and $R$ road connections, which can be represented by the traffic graph \( G = (V, E) \), where \( V = \{1, 2, \dots, N\} \) is the set of intersections and \( E \) is the set of roads. The intersections \( i, j \in V \) represent traffic intersections in the road network.

To uniquely identify each directed road, we use ordered pairs \( (i, j) \in E \), where \( 0 < i, j \leq N \) and \( i \neq j \), representing the directed road from intersection \( i \) to intersection \( j \). It is important to note that \( (i, j) \) and \( (j, i) \) represent the two different directions of the same road.        
To represent the vehicle flow conditions on real roads, we assign three types of data to each directed road \( (i,j) \): left turn probability \( \alpha_{i,j} \), right turn probability \( \beta_{i,j} \), and traffic flow \( q_{i,j} \). Here, \( \alpha_{i,j} \) and \( \beta_{i,j} \) represent the proportions of vehicles turning left and right at road \( (i,j) \) at a given moment, as a fraction of the total traffic flow \( q_{i,j} \), satisfying \( 0 \leqslant \alpha_{i,j} + \beta_{i,j} \leqslant 1 \). Naturally, the straight-through probability can be calculated as \( 1 - \alpha_{i,j} - \beta_{i,j} \).        
In the time evolution of traffic flow, the traffic flow \( q_{i,j}(t+1) \) at road \( (i,j) \) in the next time step is determined by its current traffic flow \( q_{i,j}(t) \) and the traffic flows and turning situations of all adjacent roads to \( (i,j) \). Specifically, the flow conditions on neighboring roads will influence the traffic flow changes on the current road.

To quantify and describe the traffic flow changes between two adjacent intersections, we set the east-southeast direction at \( 45^\circ \) as the positive direction of the x-axis in the Cartesian coordinate system, with the intersection point of the roads as the origin. Each road can be represented as a ray in the corresponding quadrant. Therefore, as shown in Fig. \ref{fig:TrafficFlow}, we can use superscript indices \( (1), (2), (3), (4) \) to label the roads in the four quadrants. For example, when defining the east side of intersection \( i \) connected to intersection \( j \), the left turn probability \( \alpha_{i,j} \) can be associated with the left turn probability \( \alpha_j^{(3)} \) of the road in the third quadrant of intersection \( j \), while \( \alpha_{j,i} \) is associated with the left turn probability \( \alpha_i^{(1)} \) of the road in the first quadrant of intersection \( i \).

Consider two intersections \( i \) and \( j \) connected by roads. The traffic flow difference matrices \( \mathcal{Q}_{\text{ew}} \) and \( \mathcal{Q}_{\text{sn}} \), representing the traffic flow differences when the intersections are connected in the east-west and north-south directions, respectively, can be expressed as equation \ref{eq:Q_ew} and \ref{eq:Q_sn}. 
It should be noted that in the above expression, intersection \( i \) is assumed to be located on the west or north side, and intersection \( j \) is on the east or south side. If the positions of intersections \( i \) and \( j \) are swapped, the only adjustment needed is to add a negative sign to the existing traffic flow difference, i.e., \(\mathcal{Q}_{\text{ew}} = -\mathcal{Q}_{\text{we}}\) and \(\mathcal{Q}_{\text{sn}} = -\mathcal{Q}_{\text{ns}}\).

\begin{landscape} 
\begin{align}
    \tiny
    \begin{split}
    \mathcal{Q}_{\text{ew}} = 
    & \left( \begin{array}{@{}cccc@{}}
        -q_i^{(1)}\times\alpha_i^{(1)}&0&-q_i^{(1)}\times(1-\alpha_i^{(1)})&-q_i^{(1)}\\
        -q_i^{(1)}\times\alpha_i^{(1)}&0&-q_i^{(1)}\times(1-\alpha_i^{(1)})&-q_i^{(1)}\\
        q_j^{(1)}\times(1-\alpha_j^{(1)}-\beta_j^{(1)})-q_i^{(1)}\times\alpha_i^{(1)}&q_j^{(1)}\times(1-\alpha_j^{(1)}-\beta_j^{(1)})&q_j^{(1)}\times(1-\alpha_j^{(1)}-\beta_j^{(1)})-q_i^{(1)}\times(1-\alpha_i^{(1)})&q_j^{(1)}\times(1-\alpha_j^{(1)}-\beta_j^{(1)})-q_i^{(1)}\\
        q_j^{(1)}\times(1-\alpha_j^{(1)}-\beta_j^{(1)})-q_i^{(1)}\times\alpha_i^{(1)}&q_j^{(1)}\times(1-\alpha_j^{(1)}-\beta_j^{(1)})&q_j^{(1)}\times(1-\alpha_j^{(1)}-\beta_j^{(1)})-q_i^{(1)}\times(1-\alpha_i^{(1)})&q_j^{(1)}\times(1-\alpha_j^{(1)}-\beta_j^{(1)})-q_i^{(1)}\\
        q_j^{(4)}\times\alpha_j^{(4)}-q_i^{(1)}\times\alpha_i^{(1)}&q_j^{(4)}\times\alpha_j^{(4)}&q_j^{(4)}\times\alpha_j^{(4)}-q_i^{(1)}\times(1-\alpha_i^{(1)})&q_j^{(4)}\times\alpha_j^{(4)}-q_i^{(1)}\\
        q_j^{(4)}\times\alpha_j^{(4)}-q_i^{(1)}\times\alpha_i^{(1)}&q_j^{(4)}\times\alpha_j^{(4)}&q_j^{(4)}\times\alpha_j^{(4)}-q_i^{(1)}\times(1-\alpha_i^{(1)})&q_j^{(4)}\times\alpha_j^{(4)}-q_i^{(1)}\\
        q_j^{(2)}\times\beta_j^{(2)}-q_i^{(1)}\times\alpha_i^{(1)}&q_j^{(2)}\times\beta_j^{(2)}&q_j^{(2)}\times\beta_j^{(2)}-q_i^{(1)}\times(1-\alpha_i^{(1)})&q_j^{(2)}\times\beta_j^{(2)}-q_i^{(1)}\\
        q_j^{(2)}\times\beta_j^{(2)}-q_i^{(1)}\times\alpha_i^{(1)}&q_j^{(2)}\times\beta_j^{(2)}&q_j^{(2)}\times\beta_j^{(2)}-q_i^{(1)}\times(1-\alpha_i^{(1)})&q_j^{(2)}\times\beta_j^{(2)}-q_i^{(1)}\\
    \end{array} \right. \\
    & \qquad 
    \left. \begin{array}{@{}cccc@{}}
            0&0&0&0\\
            0&0&0&0\\
            q_j^{(1)}\times(1-\alpha_j^{(1)}-\beta_j^{(1)})&q_j^{(1)}\times(1-\alpha_j^{(1)}-\beta_j^{(1)})&q_j^{(1)}\times(1-\alpha_j^{(1)}-\beta_j^{(1)})&q_j^{(1)}\times(1-\alpha_j^{(1)}-\beta_j^{(1)})\\
            q_j^{(1)}\times(1-\alpha_j^{(1)}-\beta_j^{(1)})&q_j^{(1)}\times(1-\alpha_j^{(1)}-\beta_j^{(1)})&q_j^{(1)}\times(1-\alpha_j^{(1)}-\beta_j^{(1)})&q_j^{(1)}\times(1-\alpha_j^{(1)}-\beta_j^{(1)})\\
            q_j^{(4)}\times\alpha_j^{(4)}&q_j^{(4)}\times\alpha_j^{(4)}&q_j^{(4)}\times\alpha_j^{(4)}&q_j^{(4)}\times\alpha_j^{(4)}\\
            q_j^{(4)}\times\alpha_j^{(4)}&q_j^{(4)}\times\alpha_j^{(4)}&q_j^{(4)}\times\alpha_j^{(4)}&q_j^{(4)}\times\alpha_j^{(4)}\\
            q_j^{(2)}\times\beta_j^{(2)}&q_j^{(2)}\times\beta_j^{(2)}&q_j^{(2)}\times\beta_j^{(2)}&q_j^{(2)}\times\beta_j^{(2)}\\
            q_j^{(2)}\times\beta_j^{(2)}&q_j^{(2)}\times\beta_j^{(2)}&q_j^{(2)}\times\beta_j^{(2)}&q_j^{(2)}\times\beta_j^{(2)}\\
        \end{array} \right),
    \end{split}
    \label{eq:Q_ew}
\end{align}
\begin{align}
    \tiny
    \begin{split}
    \mathcal{Q}_{\text{sn}} = 
    & \left( \begin{array}{@{}cccc@{}}
        q_j^{(3)}\times\alpha_j^{(3)} & q_j^{(3)}\times\alpha_j^{(3)} & q_j^{(3)}\times\alpha_j^{(3)} & q_j^{(3)}\times\alpha_j^{(3)}  \\
        q_j^{(3)}\times\alpha_j^{(3)} & q_j^{(3)}\times\alpha_j^{(3)} & q_j^{(3)}\times\alpha_j^{(3)} & q_j^{(3)}\times\alpha_j^{(3)} \\
        q_j^{(1)}\times\beta_j^{(1)} & q_j^{(1)}\times\beta_j^{(1)} & q_j^{(1)}\times\beta_j^{(1)} & q_j^{(1)}\times\beta_j^{(1)}  \\
        q_j^{(1)}\times\beta_j^{(1)} &  q_j^{(1)}\times\beta_j^{(1)}  & q_j^{(1)}\times\beta_j^{(1)} &  q_j^{(1)}\times\beta_j^{(1)}  \\
        0 & 0 & 0 & 0 \\
        q_j^{(4)}\times(1-\alpha_j^{(4)}-\beta_j^{(4)}) & q_j^{(4)}\times(1-\alpha_j^{(4)}-\beta_j^{(4)}) & q_j^{(4)}\times(1-\alpha_j^{(4)}-\beta_j^{(4)}) & q_j^{(4)}\times(1-\alpha_j^{(4)}-\beta_j^{(4)}) \\
        q_j^{(4)}\times(1-\alpha_j^{(4)}-\beta_j^{(4)}) & q_j^{(4)}\times(1-\alpha_j^{(4)}-\beta_j^{(4)}) & q_j^{(4)}\times(1-\alpha_j^{(4)}-\beta_j^{(4)}) & q_j^{(4)}\times(1-\alpha_j^{(4)}-\beta_j^{(4)}) \\
        0 & 0 & 0 & 0 \\
    \end{array} \right. \\
    & \qquad 
    \left. \begin{array}{@{}cccc@{}}
        q_j^{(3)}\times\alpha_j^{(3)}-q_i^{(4)}\times\alpha_i^{(4)} & q_j^{(3)}\times\alpha_j^{(3)}-q_i^{(4)} & q_j^{(3)}\times\alpha_j^{(3)}-q_i^{(4)}\times(1-\alpha_i^{(4)}) & q_j^{(3)}\times\alpha_j^{(3)} \\
        q_j^{(3)}\times\alpha_j^{(3)}-q_i^{(4)}\times\alpha_i^{(4)} & q_j^{(3)}\times\alpha_j^{(3)}-q_i^{(4)} & q_j^{(3)}\times\alpha_j^{(3)}-q_i^{(4)}\times(1-\alpha_i^{(4)}) & q_j^{(3)}\times\alpha_j^{(3)} \\
        q_j^{(1)}\times\beta_j^{(1)}-q_i^{(4)}\times\alpha_i^{(4)} & q_j^{(1)}\times\beta_j^{(1)}-q_i^{(4)} & q_j^{(1)}\times\beta_j^{(1)}-q_i^{(4)}\times(1-\alpha_i^{(4)}) & q_j^{(1)}\times\beta_j^{(1)} \\
        q_j^{(1)}\times\beta_j^{(1)}-q_i^{(4)}\times\alpha_i^{(4)} &q_j^{(1)}\times\beta_j^{(1)}-q_i^{(4)}  & q_j^{(1)}\times\beta_j^{(1)}-q_i^{(4)}\times(1-\alpha_i^{(4)}) & q_j^{(1)}\times\beta_j^{(1)} \\
        -q_i^{(4)}\times\alpha_i^{(4)} & -q_i^{(4)} & -q_i^{(4)}\times(1-\alpha_i^{(4)}) & 0 \\
        q_j^{(4)}\times(1-\alpha_j^{(4)}-\beta_j^{(4)})-q_i^{(4)}\times\alpha_i^{(4)} & q_j^{(4)}\times(1-\alpha_j^{(4)}-\beta_j^{(4)})-q_i^{(4)} & q_j^{(4)}\times(1-\alpha_j^{(4)}-\beta_j^{(4)})-q_i^{(4)}\times(1-\alpha_i^{(4)})  & q_j^{(4)}\times(1-\alpha_j^{(4)}-\beta_j^{(4)}) \\
        q_j^{(4)}\times(1-\alpha_j^{(4)}-\beta_j^{(4)})-q_i^{(4)}\times\alpha_i^{(4)} & q_j^{(4)}\times(1-\alpha_j^{(4)}-\beta_j^{(4)})-q_i^{(4)} & q_j^{(4)}\times(1-\alpha_j^{(4)}-\beta_j^{(4)})-q_i^{(4)}\times(1-\alpha_i^{(4)}) & q_j^{(4)}\times(1-\alpha_j^{(4)}-\beta_j^{(4)}) \\
        -q_i^{(4)}\times\alpha_i^{(4)} & -q_i^{(4)} & -q_i^{(4)}\times(1-\alpha_i^{(4)}) & 0 \\
        \end{array} \right).
    \end{split}
    \label{eq:Q_sn}
\end{align}
\end{landscape}

Based on \(\mathcal{Q}_{\text{ew}}\) and \(\mathcal{Q}_{\text{sn}}\), the total traffic flow entering and exiting a road at any given moment can be calculated, which allows for determining the flow variation of the road. Consequently, the flow variation equations for different roads can be derived.
First, consider the case when intersections \(i\) and \(j\) are connected in the east-west direction. According to equation (\ref{eq:Q_ew}), when the phase encoding at intersection \(j\) includes basic steering patterns XX01B, XX10B, or XX11B, there are vehicles traveling from intersection \(j\) to intersection \(i\), with traffic flow rates given by: \(q_i^{(4)}\alpha_j^{(4)}\), \(q_j^{(1)}(1-\alpha_j^{(1)}-\beta_j^{(1)})\), and \(q_j^{(2)}\beta_j^{(2)}\).        
When the phase encoding at intersection \(i\) includes the basic steering patterns 00XXB or XX10B, there are vehicles leaving the road section connecting intersections \(i\) and \(j\) through intersection \(i\), with traffic flow rates given by: \(q_i^{(1)}\alpha_i^{(1)}\) and \(q_i^{(1)}(1-\alpha_i^{(1)})\).        
Thus, the traffic flow between intersections \(j\) and \(i\) in the east-west direction can be derived as:
\begin{equation}
    \begin{split}   
        q_\text{ew}(j,i,t)=&x^{(1)}_j (1-x^{(2)}_j) q^{(4)}_j(t) \alpha^{(4)}_j
            + x^{(1)}_j x^{(2)}_j q^{(2)}_j(t) \beta^{(2)}_j
            + (1-x^{(1)}_j) x^{(2)}_j q^{(1)}_j(t) (1-\alpha^{(1)}_j-\beta^{(1)}_j)\\
            &- (1-x^{(3)}_i) (1-x^{(4)}_i) q^{(1)}_i(t) \alpha^{(1)}_i
            - x^{(2)}_i (1-x^{(1)}_i) q^{(1)}_i(t) (1-\alpha^{(1)}_i).
    \end{split}
\end{equation}

If intersections \(i\) and \(j\) are connected in the north-south direction, according to equation (\ref{eq:Q_sn}), when the phase encoding at intersection \(j\) includes the basic steering patterns XX00B, XX10B, or 11XXB, there are vehicles traveling from intersection \(j\) to intersection \(i\), with traffic flow rates given by: \(q_j^{(3)}\alpha_j^{(3)}\), \(q_j^{(1)}\alpha_j^{(1)}\), and \(q_j^{(4)}(1-\alpha_j^{(4)}-\beta_j^{(4)})\).
When the phase encoding at intersection \(i\) includes the basic steering patterns XX01B or 11XXB, there are vehicles leaving the road section connecting intersections \(i\) and \(j\) through intersection \(i\), with traffic flow rates given by: \(q_i^{(4)}\alpha_i^{(4)}\) and \(q_i^{(4)}(1-\alpha_i^{(4)})\).        
Thus, the traffic flow between intersections \(j\) and \(i\) in the north-south direction can be derived as:
\begin{equation}
    \begin{split}
        q_\text{sn}(j,i,t)=&(1-x^{(1)}_j) (1-x^{(2)}_j) q_j^{(3)}(t) \alpha_j^{(3)}
            +(1-x^{(1)}_j) x^{(2)}_j q_j^{(1)}(t) \beta_j^{(1)}
            +x^{(3)}_j x^{(4)}_j q_j^{(4)}(t) (1-\alpha_j^{(4)}-\beta_j^{(4)})\\
            &-x^{(1)}_i (1-x^{(2)}_i) q_i^{(4)}(t) \alpha_i^{(4)}
            -x^{(3)}_i x^{(4)}_i q_i^{(4)}(t) (1-\alpha_i^{(4)}).
    \end{split}
\end{equation}

Similarly, by swapping the intersection node indices \(i\) and \(j\), the flow update equations for the symmetric direction can be obtained, represented as follows:
\begin{align}
    \begin{split}
        q_\text{sn}(i,j,t)=&(1-x^{(3)}_i) (1-x^{(4)}_i) q_i^{(1)}(t) \alpha_i^{(1)}
            +(1-x^{(3)}_i) x^{(4)}_i q_i^{(3)}(t) \beta_i^{(3)}
            +x^{(1)}_i x^{(2)}_i q_i^{(2)}(t) (1-\alpha_i^{(2)}-\beta_i^{(2)})\\
            &-x^{(3)}_j (1-x^{(4)}_j) q_j^{(2)}(t) \alpha_j^{(2)}
            -x^{(1)}_j x^{(2)}_j q_j^{(2)}(t) (1-\alpha_j^{(2)}) ,    
    \end{split}\\
    \begin{split}   
        q_\text{ew}(i,j,t)=&x^{(3)}_i (1-x^{(4)}_i) q^{(2)}_i(t) \alpha^{(2)}_i
            + x^{(3)}_i x^{(4)}_i q^{(4)}_i(t) \beta^{(4)}_i
            + (1-x^{(3)}_i) x^{(4)}_i q^{(3)}_i(t) (1-\alpha^{(3)}_i-\beta^{(3)}_i)\\
            &- (1-x^{(1)}_j) (1-x^{(2)}_j) q^{(3)}_j(t) \alpha^{(3)}_j
            - x^{(4)}_j (1-x^{(3)}_j) q^{(3)}_j(t) (1-\alpha^{(3)}_j).
    \end{split}
\end{align}

Furthermore, for any road \((i,j)\), the total traffic flow at time \(t+1\) is expressed as:
\begin{equation}
    q(i,j,t+1) = 
    \begin{cases} 
        q_\text{sn}(i,j,t), & \text{if }(i,j)\text{ is south-north}, \\
        q_\text{ew}(i,j,t), & \text{if }(i,j)\text{ is east-west}.
    \end{cases}
\end{equation}

In general, for large-scale urban road networks, we assume that the more evenly distributed the vehicles are across the network, the smoother the traffic conditions. Specifically, for a given intersection, we consider that the commuting efficiency of the intersection is higher when the instantaneous traffic flow in all directions tends to be consistent. Therefore, the traffic flow deviation cost function in the network can be defined by the variance of the instantaneous traffic flow \(q(i,j,t)\) as follows:        
\begin{equation}
    H_\text{q}[\bm{x}(t), t]=\sum_{i=1}^{N}\sum_{j \in \mathcal{A}_i}\frac{{[q(j,i,t)- \overline{q}_{i}(t)]}^2}{|\mathcal{A}_i|},
\end{equation}        
where \(\mathcal{A}_i=\{j|\forall (j,i) \in E\}\) represents the set of all adjacent intersection nodes to road node \(i\), i.e., the adjacency list; \(\bm{x} \equiv [x^{(c)}_1, x^{(c)}_2, \dots, x^{(c)}_{N}]^\top\) is the state vector of the system, where \(c=1,2,3,4\). \(\overline{q}_i(t)=\sum_{j\in \mathcal{A}_i}q(j,i,t)/|\mathcal{A}_i|\) represents the average traffic flow of intersection \(i\) in each direction, and \( |\mathcal{A}_i| \equiv \text{card}{(\mathcal{A}_i)} \) represents the number of elements in \(\mathcal{A}_i\). When \(H_\text{q}(t)\to 0\), it means that the vehicle distribution in each direction of each intersection is more uniform, and the traffic is smoother at the macro level; conversely, it indicates more congestion.

Similar to the second term in equation (\ref{eq:H_ising}), to avoid frequent switching of traffic signal states and ensure smooth vehicle flow, this study introduces a signal phase change cost function \(H_{\text{d}}\), which aims to smooth the switching process of traffic signals and prevent frequent switching of signals in a short period, thereby avoiding instability in the traffic flow. Specifically,        
\begin{equation}
    H_\text{d}[\bm{x}(t),\bm{x}(t-1),t]=\eta \sum_{i=1}^{N} \sum_{c=1}^{4} D^{(c)}_i(t) ,
    \label{eq:Hd}
\end{equation}        
where \(D^{(c)}_i(t)\) represents the cost of the state change of the \(c\)-th binary bit at intersection \(i\) at time \(t\). The cost function is defined as follows:        
\begin{equation}
    D^{(c)}_i(t) =
    \begin{cases} 
        1, & \text{if } x^{(c)}_i(t-1) \neq x^{(c)}_i(t), \\
        0, & \text{if } x^{(c)}_i(t-1) = x^{(c)}_i(t).
    \end{cases}
\end{equation}        
Here, \(x^{(c)}_i(t)\) represents the phase state of the \(c\)-th binary bit at intersection \(i\) at time \(t\). The value of \(D^{(c)}_i(t)\) depends on whether the phase state at time \(t\) matches that at time \(t-1\): if they are different, \(D^{(c)}_i(t)\) is 1, indicating a state change; if the phase state is the same, \(D^{(c)}_i(t)\) is 0.         
However, according to this definition, the value of \(D^{(c)}_i(t)\) is discrete (either 0 or 1), which causes the \(H_{\text{d}}\) term in the Hamiltonian to lack a continuous derivative. Therefore, to ensure continuity and differentiability, this study proposes the following continuous approximation function to represent \(\tilde{D}^{(c)}_i(t)\):        
\begin{equation}
    \begin{split}
        \tilde{D}_i^{(c)}(t)=1-\exp\left\{-\frac{{[x_i^{(c)}(t)-x_i^{(c)}(t-1)]}^2}{2\epsilon^2}\right\}.
    \end{split}
\end{equation}        
This ensures that \(D^{(c)}_i(t)\) can take continuous values, allowing \(H_{\text{d}}\) to be differentiable. Here, \(\epsilon\) is defined as a sufficiently small positive real number. It is mathematically evident that as \(\epsilon \to 0\), \(\tilde{D}_i^{(c)}(t) = D_i^{(c)}(t)\). The parameter \(\eta\) is the weight coefficient of the \(H_{\text{d}}\) term, and its magnitude directly affects the relative penalty for traffic signal state changes compared to \(H_{\text{q}}\). In the simulation of real-world road networks, dynamic adjustment of \(\eta\) can effectively control the average duration of different phases in the network, achieving macro-level control over the speed of traffic signal switching.

Combining the previously defined \(H_q\) term, the overall Hamiltonian of the system can be expressed as:        
\begin{equation}
    H[\bm{x}(t),\bm{x}(t-1)] = H_\text{q}[\bm{x}(t)] + H_\text{d}[\bm{x}(t),\bm{x}(t-1),t].
    \label{eq:H-base}
\end{equation}

Due to the complexity of real-world road networks, further mathematical simplifications and regularizations are not performed here. However, it can be observed that for the independent variable \(\bm{x}\), since \(q_{i,j}\) is a summation of quadratic terms of \(\bm{x}\), it forms a quadratic homogeneous expression. Therefore, the Hamiltonian \(H_{\text{q}}\) contains terms of degree four in \(\bm{x}\), which is similar to the multi-spin interaction terms in the spin-glass model. On the other hand, the \(H_{\text{d}}\) term reflects the energy change caused by the variation of the state \(\bm{x}\), consisting of the summation of quadratic terms of \(\bm{x}\), similar to the second-order self-correlation terms in the spin-glass model. Thus, the model represented by equation (\ref{eq:H-base}) aligns with the typical form of a four-spin interaction model.

%% file: SupplementarySections/SS4.tex
By applying theoretical models from statistical physics to the global optimization problem of traffic signals, we can leverage existing physical optimization algorithms to solve combinatorial optimization problems, or construct corresponding physical machines to replace classical computers. This approach helps avoid the computational bottlenecks caused by the NP-hard problems faced by traditional algorithms under the von Neumann architecture \cite{chang2024quantum, Inagaki2016coherent, Mohseni2022Ising}. 
The solution time for NP-hard problems grows exponentially with the size of the problem, making it extremely difficult to solve large-scale instances. Traditional computational methods, in highly dynamic and heterogeneous environments such as urban traffic networks, often cannot effectively handle complex optimization problems due to excessive computational demands and inefficiency. 
In contrast, physical optimization algorithms and physical computers, such as quantum computing and simulated annealing, can simulate natural optimization processes, offering a potential solution. They show strong advantages, especially in solving complex systems \cite{GotoHayato2021Highperformance, SaavanPatel2020Ising}.

In physics research, solving spin-glass models with multi-spin interactions remains a significant challenge. The computational complexity of spin-glass models is high, particularly when the number of spins is large, as the model contains a large number of random couplings and local minima. This makes it difficult for traditional numerical methods to be directly applied \cite{auffinger2013random, Ros2023highdimensional}. 
Therefore, although the model theoretically has rich descriptive power, many technical challenges remain in effectively applying it to real-world problems. Currently, the academic community is exploring universal and efficient numerical methods for solving spin-glass models, such as variational methods \cite{billoire2018dynamic, franz2001exact}, Monte Carlo methods \cite{kiss2024complete, mo2023nature}, and also attempting to extend numerical algorithms developed for the Ising model, such as simulated annealing \cite{goto2019combinatorial}, quantum annealing \cite{VBapst2013Thermal, DaisukeInoue2021Traffic}, and Boltzmann machines \cite{FrancescoDAngelo2020Learning}, to solve spin-glass models.

The simulated bifurcation algorithm used in this paper is one of the methods that has shown excellent performance on the Ising model in recent years. By combining it with self-consistent field theory, we have extended it to solve spin-glass models. The simulated bifurcation algorithm gradually approaches the optimal solution through the branching behavior of physical systems. Its computational process has strong parallelism and adaptability, making it especially suitable for solving multi-spin systems. By incorporating self-consistent field theory, we can further improve the algorithm’s convergence and stability, particularly when handling many-body interactions and non-homogeneous systems.

\subsection{Simulated Bifurcation Algorithm and Improvements}

    The Simulated Bifurcation (SB) algorithm was first introduced by Goto et al. \cite{GotoHayato2021Highperformance} as a heuristic algorithm for solving combinatorial optimization problems, initially applied to the Ising model and similar binary optimization problems. The SB algorithm draws inspiration from quantum annealing and classical simulated annealing, optimizing the Hamiltonian function by simulating the bifurcation behavior of the system, gradually seeking the optimal or locally optimal solution through the physical evolution process.

    The core ideas of the SB algorithm are as follows:
    First, assume that discrete spin variables can be continuously transformed into \(\tilde{x}^{(c)}_i \in [-1, 1]\), and treat them as the generalized position coordinates in the system. This assumption converts the discrete optimization problem into a continuous optimization problem, providing a mathematical foundation for subsequent evolution.
    Second, based on the original Hamiltonian of the system, an additional self-interaction potential term \((\tilde{x}^{(c)}_i)^2\) is introduced to effectively constrain the spin change of the particles, preventing excessive response to small disturbances, thereby enhancing the algorithm's ability to search for the global optimal solution. As the algorithm iterates, this potential term gradually decreases, reducing the constraint on the particle spins, and ultimately allowing the system to find the optimal or locally optimal solution.
    Finally, the key step is the introduction of generalized momentum \(y^{(c)}_i\), which maps the variation in position space to generalized momentum space. This treatment allows for the application of classical mechanics methods, simulating the natural evolution process of the system through iterative mechanical equations.

    To apply the SB algorithm to solve the real-world road network model proposed in this paper, which uses a four-bit binary representation for each intersection's phase state, the discrete binary values \(x^{(c)}_i = 0,1\) representing each phase in equation (\ref{eq:H-base}) need to be continuously transformed. The specific mapping formula is as follows:
    \begin{equation}
        x^{(c)}_i = \frac{1 + \tilde{x}^{(c)}_i}{2} .
        \label{eq:x_smooth}
    \end{equation}
    This mapping ensures that when \(\tilde{x}^{(c)}_i \to \pm 1\), \(x^{(c)}_i\) corresponds to the binary values 0 or 1, thereby satisfying the binary description requirement for traffic phases.

    Furthermore, equation (\ref{eq:H-base}) corresponds to the general form of a four-body interaction spin-glass model, so the SB algorithm for the classical Ising model \cite{GotoHayato2021Highperformance} needs to be extended to higher dimensions. In this study, four sets of motion equations are introduced to describe the phase state of each intersection, thereby extending the SB algorithm's self-consistent iteration to a four-dimensional space. Specifically, each set of motion equations corresponds to an independent phase variable, allowing the entire intersection's state to be expressed as the dynamic evolution of a four-dimensional system. The corresponding motion equations can be written as:
    \begin{equation}
        \dot{x_i}^{(c)}=\frac{\partial H_\text{SB}}{\partial y_i^{(c)}},
        \label{eq:xic}
    \end{equation}
    \begin{equation}
        \dot{y_i}^{(c)}=-\frac{\partial H_\text{SB}}{\partial \tilde{x}_i^{(c)}}.
        \label{eq:yic}
    \end{equation}
    Where:
    \begin{equation}
        H_\text{SB}(\bm{\tilde{x}}, y)=
        \frac{a_0}{2}\sum_{c=1}^{4}\sum_{i=1}^{N}{(y_i^{(c)})}^2
        +V_\text{SB}(\bm{\tilde{x}}),
    \end{equation}
    \begin{equation}
        V_\text{SB}(\bm{\tilde{x}})=
        \sum_{c=1}^{4}\sum_{i=1}^{N}
        \left\{ \frac{1}{2} [a_0-a(\tau)] {(\tilde{x}_i^{(c)})}^2 \right\}
        + H(\bm{\tilde{x}}).
        \label{eq:V_SB}
    \end{equation}
    In the above equations, \(a_0\) and \(c_0\) are constants that control the strength of the harmonic potential \((\tilde{x}^{(c)}_i)^2\) and the external field \(H\) of the optimization objective. Typically, \(a_0 = c_0 = 1\). In this case, the bifurcation function \(a(\tau)\) is a linear function of the iteration step size, i.e., \(a(\tau) = \tau\). 
    Here, \(\tau = \Delta \tau \times Iter\), where \(\Delta \tau\) is the iteration step size, and \(Iter\) is the current iteration number, with the maximum number of iterations \(Iter_{\text{max}} = a_0 / \Delta t\). Therefore, as \(\tau\) increases, \(a(\tau)\) increases linearly until \(a_0-a(\tau)=0\), at which point the harmonic potential term loses its constraining effect, and the system converges.
    
    \begin{algorithm}
        \caption{Simulated Bifurcation Algorithm}
        \label{alg:SB}
        \begin{algorithmic}[1] 
        \STATE \textbf{Initialization:}
        \STATE Set initial state $\tilde{x}_i^{(c)}$ and initialize generalized momentum $y_i^{(c)}$ and potential $V_\text{SB}(\bm{\tilde{x}})$.
        \STATE Set algorithm parameters: initial time $t_0$, iteration step size $\Delta t$, annealing function $a(\tau)$, etc.
        
        \STATE \textbf{Iterative Evolution:}
        \REPEAT
            \STATE Update generalized momentum and position variables according to the equations of motion:
            \[
            \dot{x_i}^{(c)} = \frac{\partial H_\text{SB}}{\partial y_i^{(c)}}, \quad \dot{y_i}^{(c)} = - \frac{\partial H_\text{SB}}{\partial \tilde{x}_i^{(c)}}
            \]
            \STATE As the iteration progresses, $a(\tau)=\tau$ gradually decreases to reduce ${(\tilde{x}_i^{(c)})}^2$, allowing the system to approach an optimal or local optimal solution.
        \UNTIL{Convergence condition is met}
        
        \STATE \textbf{Termination Condition:}
        \IF{Iteration count exceeds the preset upper limit}
            \STATE Output the optimal or local optimal solution.
        \ELSE
            \STATE Continue iterating.
        \ENDIF
        \end{algorithmic}
        \label{tab:SB}
    \end{algorithm}

    In order to ensure the stability of the algorithm, the spin variable \(\tilde{x}^{(c)}_i\) is restricted to the interval \([-1, 1]\). This process can be strictly described using the following piecewise function:
    \begin{equation}
        \tilde{x}_i^{(c+1)} =
            \begin{cases}
            \tilde{x}_i^{(c+1)}, & \text{if } |\tilde{x}_i^{(c+1)}| \leq 1, \\
            \text{sgn}(\tilde{x}_i^{(c+1)}), & \text{if } |\tilde{x}_i^{(c+1)}| > 1,
            \end{cases}
    \end{equation}    
    Additionally, for the momentum variable \(y_i^{(c+1)}\), when \( |\tilde{x}_i^{(c+1)}| > 1 \), its value is forced to be 0:    
    \begin{equation}
        y_i^{(c+1)} =
            \begin{cases}
            y_i^{(c+1)}, & \text{if } |\tilde{x}_i^{(c+1)}| \leq 1, \\
            0, & \text{if } |\tilde{x}_i^{(c+1)}| > 1.
            \end{cases}
    \end{equation}    
    Here, \(\text{sgn}(\tilde{x}_i)\) is the sign function, defined as:
    \begin{equation}
        \text{sgn}(\tilde{x}_i) =
            \begin{cases}
            1, & \text{if } \tilde{x}_i > 0, \\
            -1, & \text{if } \tilde{x}_i < 0, \\
            0, & \text{if } \tilde{x}_i = 0.
            \end{cases}
    \end{equation}     
    This correction method ensures that \(\tilde{x}_i\) always stays within the interval \([-1, 1]\) and applies the necessary constraint on the momentum variable \(y_i\) when \(\tilde{x}_i\) exceeds this interval. The basic flow of the SB algorithm is presented in Algorithm (\ref{tab:SB}).

    \begin{figure}[tpb]
        \centering
        \includegraphics[width=\textwidth]{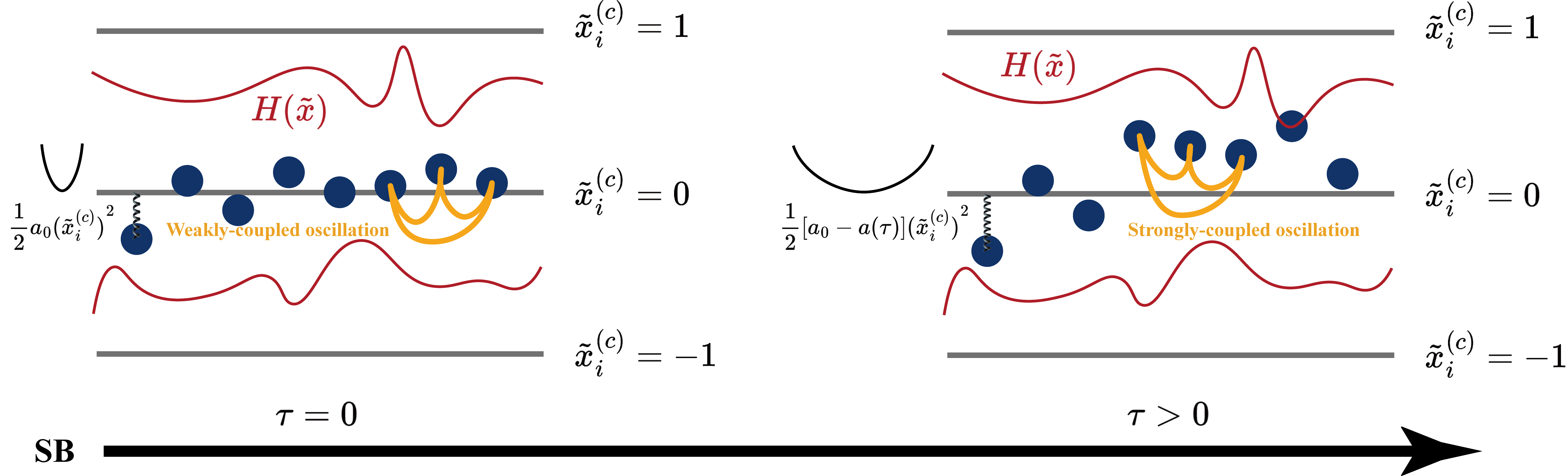}
        \caption{Schematic of the SB algorithm convergence process. The red line represents the $H$ surface of the system, which corresponds to the cost function we aim to minimize. As the simulation time $\tau$ increases, the constraint of the elastic potential well gradually weakens, causing each spin variable to deviate further from zero and become increasingly influenced by interactions, transitioning from weakly-coupled oscillation to strongly-coupled oscillation.}
        \label{fig:SimulatedBifurcation}
    \end{figure}

    As shown in Fig. \ref{fig:SimulatedBifurcation}, the physical picture of the SB algorithm can be described as follows: each spin variable \(x^{(c)}_i\) to be solved can be treated as a classical particle after transformation by equation (\ref{eq:x_smooth}), with generalized position \(\tilde{x}^{(c)}_i\) and generalized momentum \(y^{(c)}_i\). The generalized position vector \(\bm{\tilde{x}} \equiv \{ \tilde{x}^{(c)}_i \}\) and the momentum field \(\bm{y}=\{ y^{(c)}_i \}\) together define a state space that satisfies the uniformity assumption in theoretical mechanics. Each particle \(\tilde{x}^{(c)}_i\) is constrained by an additional harmonic potential \((\tilde{x}^{(c)}_i)^2\), limiting its position near the origin (i.e., \(\tilde{x}^{(c)}_i = 0\)).
    At the system's initialization stage, in order to break the symmetry, we apply small perturbations to each particle's position so that all particles oscillate slightly around the potential minimum of the harmonic potential. If the interactions between particles are not considered, as the iteration progresses, the harmonic potential's constraint gradually weakens, and the particles randomly drift towards either \(+1\) or \(-1\).
    
    However, when considering particle interactions, the system becomes more complex. The interactions between particles effectively establish an additional spring network between the particles oscillating near their equilibrium positions. As a result, the motion of each particle is no longer independent but is influenced by the generalized position \(\tilde{x}^{(c)}_j\) of the particles it interacts with. For example, when a particle's motion reaches its maximum amplitude, it may exactly satisfy the potential constraint of neighboring particles, thus finding a state with minimum potential energy. At this point, the particle, along with the associated particles, forms a coupled oscillation group, maintaining this low-energy state as much as possible until the system converges to a minimum value of the Hamiltonian or reaches a state with a lower Hamiltonian.
    
    Thanks to the gradual weakening of the harmonic potential constraint, the SB algorithm demonstrates better global convergence. Specifically, the gradual increase of \(a(\tau)\) means that the external field’s influence is introduced progressively, allowing the system to smoothly transition between different local optima. Meanwhile, the harmonic potential causes each particle’s generalized position \(\tilde{x}^{(c)}_j\) to continuously oscillate between positive and negative values. This dynamic oscillation effectively helps the system cross energy barriers between local optima, avoiding being trapped in local minima. Through this mechanism, the system can gradually evolve towards the global optimum, ensuring better global convergence.
    
    The computational advantage of the SB algorithm lies in its optimization based on field theory, avoiding the complexity of traditional numerical optimization methods, which require substantial multi-particle interactions. Instead, it approximates the interactions between particles as the interaction between a single particle and its conjugate field. This idea is reflected in the definition of the generalized momentum \(y^{(c)}_i\), which plays a key auxiliary role in the continuous evolution of the mechanical process.    
    It is important to note that the calculation of \(y^{(c)}_i\) considers the influence of particle interactions, while the update of the generalized position depends only on the particle’s own momentum and is independent of multi-body interactions between other particles, as shown in equations (\ref{eq:xic}) and (\ref{eq:yic}). This decouples the particles, meaning that the positions can be updated directly based on the momentum field, avoiding the extra computational burden and time complexity of recalculating Hamiltonian changes when updating particle positions, as in other numerical optimization algorithms. Additionally, the decoupling of particles allows the algorithm to be efficiently distributed across multiple computational cores, enabling high-performance parallel computation.
    
    In the actual problem-solving process, attention must also be given to potential additional constraints. For example, in the global optimization problem for traffic lights in this study, there are nonlinear constraints between the four-bit binary numbers in equation (\ref{eq:H-base}). For instance, \(0001\text{B}\) is an invalid state, as it cannot encode any valid traffic signal among the eight known phases. Thus, the traffic network optimization problem is transformed into a problem of solving partial differential equations under constraints. Specifically, among the possible four sets of binary numbers, only the eight states shown in Fig. 1(b) in the main text are valid traffic light phases, denoted as the allowed set \(M = \{0000\text{B}, 0010\text{B}, 0111\text{B}, 1010\text{B}, 0101\text{B}, 1000\text{B}, 1101\text{B}, 1111\text{B}\}\). The remaining eight invalid states form the forbidden set \(\varLambda = \{0001\text{B}, 0011\text{B}, 0110\text{B}, 0100\text{B}, 1001\text{B}, 1011\text{B}, 1100\text{B}, 1110\text{B}\}\).    
    Therefore, during the solution process, a Lagrange function \(H_w\) needs to be introduced to restrict the system's free evolution, ensuring that the search occurs only within the allowed state space. To guarantee the effectiveness of the SB algorithm, \(H_w\) should be a continuous function and be continuously differentiable within the domain of \(\tilde{x}^{(c)}_i\). A single-node constraint function can be defined as follows:    
    \begin{equation}
        g(x,m) = \sum_{c=1}^{4}| x^{(c)}-m^{(c)} |.
    \end{equation}    
    This function represents that when the current state \(x\) exactly matches the target state \(m\) (i.e., when the four-bit binary numbers match), \(g(x,m) = 0\); otherwise, \(g(x,m)\) is nonzero, resulting in an additional energy cost.

    \( H_\text{w} \) can be defined as follows:
    \begin{equation}
        H_\text{w}(\bm{\tilde{x}}) = \zeta \sum_{i=1}^{N} \sum_{c=1}^{4} 
        G(\tilde{x}^{(c)}_i, m),
    \end{equation}
    \begin{equation}
        G(\tilde{x}^{(c)}_i, m) =
        \prod_{m \in M} g(\tilde{x}^{(c)}_i, m),
    \end{equation}
    where \(\zeta\) is the Lagrange multiplier, typically set as a large constant to control the strictness of the constraint. 
    If the intersection phase encoding belongs to any of the eight allowed four-bit binary numbers, then \( H_\text{w} = 0 \); otherwise, \( H_\text{w} \) will contribute to the Hamiltonian. 
    It is important to note that in the above constraints, for all the binary numbers in the set \(M\), the binary bits with value 0 are replaced by -1 to meet the requirements of the SB algorithm.

    Therefore, equation (\ref{eq:H-base}) needs to be rewritten as:
    \begin{equation}
        H[\bm{x}(t),\bm{x}(t-1)] = H_\text{q}[\bm{x}(t)] + H_\text{d}[\bm{x}(t),\bm{x}(t-1),t] + H_\text{w}[\bm{\tilde{x}}(t)],
        \label{eq:H-base-fix}
    \end{equation}    
    Then, it is substituted into equation (\ref{eq:V_SB}) to obtain \(H_{\text{SB}}\).    
    To ensure the stability of the system, an alternating implicit scheme is used in the numerical solution of the SB algorithm, with the Hamiltonian equations being solved numerically in the following order:    
    \begin{equation}
        \dot{y}^{(c)}_i=-\frac{\partial H_{\text{SB}}}{\partial \tilde{x}_i^{(c)}},
    \end{equation}    
    \begin{equation}
        \dot{x}^{(c)}_i=\frac{\partial H_{\text{SB}}}{\partial y_i^{(c)}},
    \end{equation}    

    Substituting into equation (\ref{eq:H-base-fix}), the expressions for the derivatives in the above equations can be given.
    \begin{equation}
        \dot{x}^{(c)}_i=\frac{\partial H_{\mathrm{SB}}}{\partial y_i^{(c)}}=a_0y_i^{(c)},
    \end{equation}
    \begin{equation}
            \dot{y}_i^{(c)}
            =-\frac{\partial H_{\mathrm{SB}}}{\partial \tilde{x}_i^{(c)}}
            =-(a_0-a(\tau))\tilde{x}_i^{(c)}
            -\frac{\partial H_{\mathrm{q}}}{\partial \tilde{x}_i^{(c)}}
            -\frac{\partial H_{\mathrm{d}}}{\partial \tilde{x}_i^{(c)}}
            -\frac{\partial H_{\mathrm{w}}}{\partial \tilde{x}_i^{(c)}},
    \end{equation}
    where:
    \begin{equation}
        \frac{\partial H_{\text{q}}}{\partial \tilde{x}^{(c)}_i} = 
        \sum_{j \in \mathcal{A}_i}
        \frac{2}{|\mathcal{A}_i|} (q_{j,i}-\bar{q}_i)
        (\frac{\partial q_{j,i}}{\partial \tilde{x}^{(c)}_i}  
            - \frac{\partial \bar{q}_i}{\partial \tilde{x}^{(c)}_i} )
        + \sum_{j \in \mathcal{A}_i}
        \frac{2}{|\mathcal{A}_j|} (q_{i,j}-\bar{q}_j)
        \frac{|\mathcal{A}_j| - 1}{|\mathcal{A}_j|} 
        \frac{\partial q_{i,j}}{\partial \tilde{x}^{(c)}_i} ,
    \end{equation}
    \begin{equation}
        \frac{\partial H_{\text{d}}}{\partial \tilde{x}^{(c)}_i} = 
        \frac{1}{\epsilon ^2} (\tilde{x}_i^{(c)}-\mu_i^{(c)})
        \exp\left\{-\frac{{(\tilde{x}_i^{(c)}-\mu_i^{(c)})}^2}{2\epsilon^2}\right\},
    \end{equation}
    \begin{equation}
        \frac{\partial H_{\text{w}}}{\partial \tilde{x}^{(c)}_i} = 
        \zeta \frac{\partial G(\tilde{x}^{(c)}_i, m)}{\partial \tilde{x}^{(c)}_i} = 
        \zeta \sum_{m\in M}\frac{G(\tilde{x}^{(c)}_i, m)}{g(x_i^{(c)},m)}.
    \end{equation}

\subsection{Baseline: Simulated annealing algorithm}

    To demonstrate the computational advantages of the SB algorithm proposed in this paper, we choose the simulated annealing method as a comparison.
    Simulated annealing (SA) is a probabilistic optimization algorithm used to find the optimal solution in large-scale search spaces. Its principle is inspired by the annealing process in thermodynamics, where the system's temperature is gradually lowered, allowing the system to evolve from a high-energy state to a low-energy state. This process helps avoid the system getting trapped in local minima to some extent \cite{kirkpatrick1983optimization}. 
    In SA, the algorithm starts from an initial solution, selects a neighboring solution at each step, and calculates the Hamiltonian difference \(\Delta H\) between the current solution and the new one. According to the Metropolis criterion \cite{delahaye2019simulated}, the probability of accepting the new solution is given by:
    \begin{equation}
        P_{r}(\text{Current state} = j) = \exp\left( \frac{E_i - E_j}{k_{\text{B}} T}\right),
    \end{equation}
    where \(E_i\) and \(E_j\) are the energies of the current and new states, \(k_{\text{B}}\) is the Boltzmann constant, and \(T\) is the current temperature.

    As the temperature is gradually lowered, the algorithm transitions from higher energy states to lower energy states, ultimately finding the global optimum. The temperature update formula is:
    \begin{equation}
        T_{\text{new}} = T_{\text{old}} \times \omega ,
    \end{equation}
    where \(\omega\) is the annealing rate, typically a value between \(0\) and \(1\). As the temperature decreases, the disturbance amplitude of the system gradually reduces, eventually converging to either a local or global optimal solution.

%% file: main.bbl

\begin{thebibliography}{76}
\ifx \bisbn   \undefined \def \bisbn  #1{ISBN #1}\fi
\ifx \binits  \undefined \def \binits#1{#1}\fi
\ifx \bauthor  \undefined \def \bauthor#1{#1}\fi
\ifx \batitle  \undefined \def \batitle#1{#1}\fi
\ifx \bjtitle  \undefined \def \bjtitle#1{#1}\fi
\ifx \bvolume  \undefined \def \bvolume#1{\textbf{#1}}\fi
\ifx \byear  \undefined \def \byear#1{#1}\fi
\ifx \bissue  \undefined \def \bissue#1{#1}\fi
\ifx \bfpage  \undefined \def \bfpage#1{#1}\fi
\ifx \blpage  \undefined \def \blpage #1{#1}\fi
\ifx \burl  \undefined \def \burl#1{\textsf{#1}}\fi
\ifx \doiurl  \undefined \def \doiurl#1{\url{https://doi.org/#1}}\fi
\ifx \betal  \undefined \def \betal{\textit{et al.}}\fi
\ifx \binstitute  \undefined \def \binstitute#1{#1}\fi
\ifx \binstitutionaled  \undefined \def \binstitutionaled#1{#1}\fi
\ifx \bctitle  \undefined \def \bctitle#1{#1}\fi
\ifx \beditor  \undefined \def \beditor#1{#1}\fi
\ifx \bpublisher  \undefined \def \bpublisher#1{#1}\fi
\ifx \bbtitle  \undefined \def \bbtitle#1{#1}\fi
\ifx \bedition  \undefined \def \bedition#1{#1}\fi
\ifx \bseriesno  \undefined \def \bseriesno#1{#1}\fi
\ifx \blocation  \undefined \def \blocation#1{#1}\fi
\ifx \bsertitle  \undefined \def \bsertitle#1{#1}\fi
\ifx \bsnm \undefined \def \bsnm#1{#1}\fi
\ifx \bsuffix \undefined \def \bsuffix#1{#1}\fi
\ifx \bparticle \undefined \def \bparticle#1{#1}\fi
\ifx \barticle \undefined \def \barticle#1{#1}\fi
\bibcommenthead
\ifx \bconfdate \undefined \def \bconfdate #1{#1}\fi
\ifx \botherref \undefined \def \botherref #1{#1}\fi
\ifx \url \undefined \def \url#1{\textsf{#1}}\fi
\ifx \bchapter \undefined \def \bchapter#1{#1}\fi
\ifx \bbook \undefined \def \bbook#1{#1}\fi
\ifx \bcomment \undefined \def \bcomment#1{#1}\fi
\ifx \oauthor \undefined \def \oauthor#1{#1}\fi
\ifx \citeauthoryear \undefined \def \citeauthoryear#1{#1}\fi
\ifx \endbibitem  \undefined \def \endbibitem {}\fi
\ifx \bconflocation  \undefined \def \bconflocation#1{#1}\fi
\ifx \arxivurl  \undefined \def \arxivurl#1{\textsf{#1}}\fi
\csname PreBibitemsHook\endcsname

\bibitem[\protect\citeauthoryear{{E. A. Stanciu}
  et~al.}{0025/2012-10-27}]{E.A.Stanciu25Optimization}
\begin{bchapter}
\bauthor{\bsnm{{E. A. Stanciu}}},
\bauthor{\bsnm{{I. M. Moise}}},
\bauthor{\bsnm{{L. M. Nemtoi}}}:
\bctitle{Optimization of urban road traffic in {{Intelligent Transport
  Systems}}}.
In: \bbtitle{2012 {{International Conference}} on {{Applied}} and {{Theoretical
  Electricity}} ({{ICATE}})},
pp. \bfpage{1}--\blpage{4}
(\byear{0025/2012-10-27}).
\doiurl{10.1109/ICATE.2012.6403458}
\end{bchapter}
\endbibitem

\bibitem[\protect\citeauthoryear{{L. Butler} et~al.}{2020}]{L.Butler2020Smart}
\begin{barticle}
\bauthor{\bsnm{{L. Butler}}},
\bauthor{\bsnm{{T. Yigitcanlar}}},
\bauthor{\bsnm{{A. Paz}}}:
\batitle{Smart {{Urban Mobility Innovations}}: {{A Comprehensive Review}} and
  {{Evaluation}}}.
\bjtitle{IEEE Access}
\bvolume{8},
\bfpage{196034}--\blpage{196049}
(\byear{2020})
\doiurl{10.1109/ACCESS.2020.3034596}
\end{barticle}
\endbibitem

\bibitem[\protect\citeauthoryear{Zhu et~al.}{2020}]{Zhu2020Parallel}
\begin{barticle}
\bauthor{\bsnm{Zhu}, \binits{F.}},
\bauthor{\bsnm{Lv}, \binits{Y.}},
\bauthor{\bsnm{Chen}, \binits{Y.}},
\bauthor{\bsnm{Wang}, \binits{X.}},
\bauthor{\bsnm{Xiong}, \binits{G.}},
\bauthor{\bsnm{Wang}, \binits{F.-Y.}}:
\batitle{Parallel transportation systems: {{Toward IoT-enabled}} smart urban
  traffic control and management}.
\bjtitle{IEEE Transactions on Intelligent Transportation Systems}
\bvolume{21}(\bissue{10}),
\bfpage{4063}--\blpage{4071}
(\byear{2020})
\doiurl{10.1109/TITS.2019.2934991}
\end{barticle}
\endbibitem

\bibitem[\protect\citeauthoryear{Vitk{\=u}nas
  et~al.}{2021}]{Vitkunas2021Assessment}
\begin{barticle}
\bauthor{\bsnm{Vitk{\=u}nas}, \binits{R.}},
\bauthor{\bsnm{{\v C}in{\v c}ikait{\.e}}, \binits{R.}},
\bauthor{\bsnm{{Meidute-Kavaliauskiene}}, \binits{I.}}:
\batitle{Assessment of the {{Impact}} of {{Road Transport Change}} on the
  {{Security}} of the {{Urban Social Environment}}}.
\bjtitle{Sustainability}
\bvolume{13}(\bissue{22}),
\bfpage{12630}
(\byear{2021})
\doiurl{10.3390/su132212630}
\end{barticle}
\endbibitem

\bibitem[\protect\citeauthoryear{Bharadiya}{2023}]{Bharadiya2023Artificial}
\begin{barticle}
\bauthor{\bsnm{Bharadiya}, \binits{J.}}:
\batitle{Artificial {{Intelligence}} in {{Transportation Systems A Critical
  Review}}}.
\bjtitle{American Journal of Computing and Engineering}
\bvolume{6}(\bissue{1}),
\bfpage{34}--\blpage{45}
(\byear{2023})
\doiurl{10.47672/ajce.1487}
\end{barticle}
\endbibitem

\bibitem[\protect\citeauthoryear{Nigam et~al.}{2023}]{Nigam2023Review}
\begin{barticle}
\bauthor{\bsnm{Nigam}, \binits{N.}},
\bauthor{\bsnm{Singh}, \binits{D.P.}},
\bauthor{\bsnm{Choudhary}, \binits{J.}}:
\batitle{A {{Review}} of {{Different Components}} of the {{Intelligent Traffic
  Management System}} ({{ITMS}})}.
\bjtitle{Symmetry}
\bvolume{15}(\bissue{3}),
\bfpage{583}
(\byear{2023})
\doiurl{10.3390/sym15030583}
\end{barticle}
\endbibitem

\bibitem[\protect\citeauthoryear{Samaei}{2024}]{Samaei2024Using}
\begin{bchapter}
\bauthor{\bsnm{Samaei}, \binits{S.R.}}:
\bctitle{Using artificial intelligence to increase urban resilience: A case
  study of {{Tehran}}}.
In: \bbtitle{13th {{International Conference}} on {{Advanced Research}} in
  {{Science}}, {{Engineering}} and {{Technology}}, {{Brussels}}, {{Belgium}}}
(\byear{2024})
\end{bchapter}
\endbibitem

\bibitem[\protect\citeauthoryear{Hao and Boel}{2022}]{Hao2022Convergence}
\begin{barticle}
\bauthor{\bsnm{Hao}, \binits{Z.}},
\bauthor{\bsnm{Boel}, \binits{R.}}:
\batitle{Convergence analysis on control for traffic signals in urban road
  network}.
\bjtitle{Transportation Research Part B: Methodological}
\bvolume{165},
\bfpage{35}--\blpage{62}
(\byear{2022})
\doiurl{10.1016/j.trb.2022.09.006}
\end{barticle}
\endbibitem

\bibitem[\protect\citeauthoryear{Yu et~al.}{2018}]{Yu2018Integrated}
\begin{barticle}
\bauthor{\bsnm{Yu}, \binits{C.}},
\bauthor{\bsnm{Feng}, \binits{Y.}},
\bauthor{\bsnm{Liu}, \binits{H.X.}},
\bauthor{\bsnm{Ma}, \binits{W.}},
\bauthor{\bsnm{Yang}, \binits{X.}}:
\batitle{Integrated optimization of traffic signals and vehicle trajectories at
  isolated urban intersections}.
\bjtitle{Transportation Research Part B: Methodological}
\bvolume{112},
\bfpage{89}--\blpage{112}
(\byear{2018})
\doiurl{10.1016/j.trb.2018.04.007}
\end{barticle}
\endbibitem

\bibitem[\protect\citeauthoryear{Webster}{1958}]{Webster1958Traffic}
\begin{botherref}
\oauthor{\bsnm{Webster}, \binits{F.V.}}:
Traffic {{Signal Settings}}.
Road Research Technical Paper
(1958)
\end{botherref}
\endbibitem

\bibitem[\protect\citeauthoryear{Lowrie}{1990}]{Lowrie1990Scats}
\begin{botherref}
\oauthor{\bsnm{Lowrie}, \binits{P.R.}}:
Scats, sydney co-ordinated adaptive traffic system: {{A}} traffic responsive
  method of controlling urban traffic
(1990)
\end{botherref}
\endbibitem

\bibitem[\protect\citeauthoryear{Hunt et~al.}{1981}]{Hunt1981SCOOTaTR}
\begin{bchapter}
\bauthor{\bsnm{Hunt}, \binits{P.B.}},
\bauthor{\bsnm{Robertson}, \binits{D.I.}},
\bauthor{\bsnm{Bretherton}, \binits{R.D.}},
\bauthor{\bsnm{Winton}, \binits{R.I.}}:
\bctitle{{{SCOOT-a}} traffic responsive method of coordinating signals}.
(\byear{1981})
\end{bchapter}
\endbibitem

\bibitem[\protect\citeauthoryear{{Cabrejas-Egea}
  et~al.}{2021}]{Cabrejas-Egea2021Reinforcement}
\begin{barticle}
\bauthor{\bsnm{{Cabrejas-Egea}}, \binits{A.}},
\bauthor{\bsnm{Zhang}, \binits{R.}},
\bauthor{\bsnm{Walton}, \binits{N.}}:
\batitle{Reinforcement learning for traffic signal control: {{Comparison}} with
  commercial systems}.
\bjtitle{Transportation research procedia}
\bvolume{58},
\bfpage{638}--\blpage{645}
(\byear{2021})
\end{barticle}
\endbibitem

\bibitem[\protect\citeauthoryear{Du et~al.}{2023}]{Du2023Safelight}
\begin{bchapter}
\bauthor{\bsnm{Du}, \binits{W.}},
\bauthor{\bsnm{Ye}, \binits{J.}},
\bauthor{\bsnm{Gu}, \binits{J.}},
\bauthor{\bsnm{Li}, \binits{J.}},
\bauthor{\bsnm{Wei}, \binits{H.}},
\bauthor{\bsnm{Wang}, \binits{G.}}:
\bctitle{Safelight: {{A}} reinforcement learning method toward collision-free
  traffic signal control}.
In: \bbtitle{Proceedings of the {{AAAI}} Conference on Artificial
  Intelligence},
vol. \bseriesno{37},
pp. \bfpage{14801}--\blpage{14810}
(\byear{2023})
\end{bchapter}
\endbibitem

\bibitem[\protect\citeauthoryear{Bi et~al.}{2014}]{Bi2014Type2}
\begin{barticle}
\bauthor{\bsnm{Bi}, \binits{Y.}},
\bauthor{\bsnm{Srinivasan}, \binits{D.}},
\bauthor{\bsnm{Lu}, \binits{X.}},
\bauthor{\bsnm{Sun}, \binits{Z.}},
\bauthor{\bsnm{Zeng}, \binits{W.}}:
\batitle{Type-2 fuzzy multi-intersection traffic signal control with
  differential evolution optimization}.
\bjtitle{Expert systems with applications}
\bvolume{41}(\bissue{16}),
\bfpage{7338}--\blpage{7349}
(\byear{2014})
\end{barticle}
\endbibitem

\bibitem[\protect\citeauthoryear{Nair and Cai}{2007}]{Nair2007fuzzy}
\begin{bchapter}
\bauthor{\bsnm{Nair}, \binits{B.M.}},
\bauthor{\bsnm{Cai}, \binits{J.}}:
\bctitle{A fuzzy logic controller for isolated signalized intersection with
  traffic abnormality considered}.
In: \bbtitle{2007 {{IEEE}} Intelligent Vehicles Symposium},
pp. \bfpage{1229}--\blpage{1233}.
\bpublisher{IEEE},
\blocation{New York, NY}
(\byear{2007})
\end{bchapter}
\endbibitem

\bibitem[\protect\citeauthoryear{Zhang et~al.}{2017}]{Zhang2017Application}
\begin{barticle}
\bauthor{\bsnm{Zhang}, \binits{N.}},
\bauthor{\bsnm{Zhou}, \binits{K.}},
\bauthor{\bsnm{Du}, \binits{X.}}:
\batitle{Application of fuzzy logic and fuzzy {{AHP}} to mineral prospectivity
  mapping of porphyry and hydrothermal vein copper deposits in the
  {{Dananhu-Tousuquan}} island arc, {{Xinjiang}}, {{NW China}}}.
\bjtitle{Journal of African Earth Sciences}
\bvolume{128},
\bfpage{84}--\blpage{96}
(\byear{2017})
\end{barticle}
\endbibitem

\bibitem[\protect\citeauthoryear{Yu et~al.}{2018}]{Yu2018Optimal}
\begin{barticle}
\bauthor{\bsnm{Yu}, \binits{H.}},
\bauthor{\bsnm{Ma}, \binits{R.}},
\bauthor{\bsnm{Zhang}, \binits{H.M.}}:
\batitle{Optimal traffic signal control under dynamic user equilibrium and link
  constraints in a general network}.
\bjtitle{Transportation Research Part B: Methodological}
\bvolume{110},
\bfpage{302}--\blpage{325}
(\byear{2018})
\doiurl{10.1016/j.trb.2018.02.009}
\end{barticle}
\endbibitem

\bibitem[\protect\citeauthoryear{Qadri et~al.}{2020}]{Qadri2020Stateofart}
\begin{barticle}
\bauthor{\bsnm{Qadri}, \binits{S.S.S.M.}},
\bauthor{\bsnm{G{\"o}k{\c c}e}, \binits{M.A.}},
\bauthor{\bsnm{{\"O}ner}, \binits{E.}}:
\batitle{State-of-art review of traffic signal control methods: Challenges and
  opportunities}.
\bjtitle{European transport research review}
\bvolume{12},
\bfpage{1}--\blpage{23}
(\byear{2020})
\end{barticle}
\endbibitem

\bibitem[\protect\citeauthoryear{PAIK and SAHNI}{1995}]{PAIK1995NETWORK}
\begin{barticle}
\bauthor{\bsnm{PAIK}, \binits{D.}},
\bauthor{\bsnm{SAHNI}, \binits{S.}}:
\batitle{{{NETWORK UPGRADING PROBLEMS}}}.
\bjtitle{NETWORKS}
\bvolume{26}(\bissue{1}),
\bfpage{45}--\blpage{58}
(\byear{1995})
\doiurl{10.1002/net.3230260105}
\end{barticle}
\endbibitem

\bibitem[\protect\citeauthoryear{Renfrew}{2009}]{Renfrew2009Traffic}
\begin{bbook}
\bauthor{\bsnm{Renfrew}, \binits{D.}}:
\bbtitle{Traffic {{Signal Control}} with {{Ant Colony Optimization}}},
(\byear{2009})
\end{bbook}
\endbibitem

\bibitem[\protect\citeauthoryear{Tchuitcheu
  et~al.}{2020}]{Tchuitcheu2020Internet}
\begin{botherref}
\oauthor{\bsnm{Tchuitcheu}, \binits{W.C.}},
\oauthor{\bsnm{Bobda}, \binits{C.}},
\oauthor{\bsnm{Pantho}, \binits{M.J.H.}}:
Internet of smart-cameras for traffic lights optimization in smart cities.
INTERNET OF THINGS
\textbf{11}
(2020)
\doiurl{10.1016/j.iot.2020.100207}
\end{botherref}
\endbibitem

\bibitem[\protect\citeauthoryear{Chu et~al.}{2019}]{Chu2019Multiagent}
\begin{barticle}
\bauthor{\bsnm{Chu}, \binits{T.}},
\bauthor{\bsnm{Wang}, \binits{J.}},
\bauthor{\bsnm{Codec{\`a}}, \binits{L.}},
\bauthor{\bsnm{Li}, \binits{Z.}}:
\batitle{Multi-agent deep reinforcement learning for large-scale traffic signal
  control}.
\bjtitle{IEEE transactions on intelligent transportation systems}
\bvolume{21}(\bissue{3}),
\bfpage{1086}--\blpage{1095}
(\byear{2019})
\end{barticle}
\endbibitem

\bibitem[\protect\citeauthoryear{{El-Tantawy} and
  Abdulhai}{2010}]{El-Tantawy2010agentbased}
\begin{bchapter}
\bauthor{\bsnm{{El-Tantawy}}, \binits{S.}},
\bauthor{\bsnm{Abdulhai}, \binits{B.}}:
\bctitle{An agent-based learning towards decentralized and coordinated traffic
  signal control}.
In: \bbtitle{13th {{International IEEE}} Conference on Intelligent
  Transportation Systems},
pp. \bfpage{665}--\blpage{670}.
\bpublisher{IEEE},
\blocation{New York, NY}
(\byear{2010})
\end{bchapter}
\endbibitem

\bibitem[\protect\citeauthoryear{Ge et~al.}{2014}]{Ge2014EnergySustainable}
\begin{barticle}
\bauthor{\bsnm{Ge}, \binits{X.-Y.}},
\bauthor{\bsnm{Li}, \binits{Z.-C.}},
\bauthor{\bsnm{Lam}, \binits{W.H.K.}},
\bauthor{\bsnm{Choi}, \binits{K.}}:
\batitle{Energy-{{Sustainable Traffic Signal Timings}} for a {{Congested Road
  Network With Heterogeneous Users}}}.
\bjtitle{IEEE TRANSACTIONS ON INTELLIGENT TRANSPORTATION SYSTEMS}
\bvolume{15}(\bissue{3}),
\bfpage{1016}--\blpage{1025}
(\byear{2014})
\doiurl{10.1109/TITS.2013.2291612}
\end{barticle}
\endbibitem

\bibitem[\protect\citeauthoryear{Renfrew et~al.}{2012}]{Renfrew2012Traffic}
\begin{bchapter}
\bauthor{\bsnm{Renfrew}, \binits{D.}},
\bauthor{\bsnm{Yu}, \binits{X.-H.}},
\bauthor{\bsnm{{IEEE}}}:
\bctitle{Traffic {{Signal Optimization Using Ant Colony Algorithm}}}.
In: \bbtitle{2012 {{INTERNATIONAL JOINT CONFERENCE ON NEURAL NETWORKS}}
  ({{IJCNN}})}
(\byear{2012})
\end{bchapter}
\endbibitem

\bibitem[\protect\citeauthoryear{Eom and Kim}{2020}]{Eom2020traffic}
\begin{barticle}
\bauthor{\bsnm{Eom}, \binits{M.}},
\bauthor{\bsnm{Kim}, \binits{B.-I.}}:
\batitle{The traffic signal control problem for intersections: A review}.
\bjtitle{European transport research review}
\bvolume{12},
\bfpage{1}--\blpage{20}
(\byear{2020})
\end{barticle}
\endbibitem

\bibitem[\protect\citeauthoryear{{Daisuke
  Inoue}}{2021}]{DaisukeInoue2021Traffic}
\begin{botherref}
\oauthor{\bsnm{{Daisuke Inoue}}}:
Traffic signal optimization on a square lattice with quantum annealing.
Scientific Reports,
12
(2021)
\end{botherref}
\endbibitem

\bibitem[\protect\citeauthoryear{Chang et~al.}{2024}]{chang2024quantum}
\begin{botherref}
\oauthor{\bsnm{Chang}, \binits{Y.-J.}},
\oauthor{\bsnm{Nien}, \binits{C.-F.}},
\oauthor{\bsnm{Huang}, \binits{K.-P.}},
\oauthor{\bsnm{Zhang}, \binits{Y.-T.}},
\oauthor{\bsnm{Cho}, \binits{C.-H.}},
\oauthor{\bsnm{Chang}, \binits{C.-R.}}:
Quantum computing for optimization with ising machine.
IEEE Nanotechnology Magazine
(2024)
\end{botherref}
\endbibitem

\bibitem[\protect\citeauthoryear{Inagaki et~al.}{2016}]{Inagaki2016coherent}
\begin{barticle}
\bauthor{\bsnm{Inagaki}, \binits{T.}},
\bauthor{\bsnm{Haribara}, \binits{Y.}},
\bauthor{\bsnm{Igarashi}, \binits{K.}},
\bauthor{\bsnm{Sonobe}, \binits{T.}},
\bauthor{\bsnm{Tamate}, \binits{S.}},
\bauthor{\bsnm{Honjo}, \binits{T.}},
\bauthor{\bsnm{Marandi}, \binits{A.}},
\bauthor{\bsnm{McMahon}, \binits{P.L.}},
\bauthor{\bsnm{Umeki}, \binits{T.}},
\bauthor{\bsnm{Enbutsu}, \binits{K.}}:
\batitle{A coherent {{Ising}} machine for 2000-node optimization problems}.
\bjtitle{Science}
\bvolume{354}(\bissue{6312}),
\bfpage{603}--\blpage{606}
(\byear{2016})
\end{barticle}
\endbibitem

\bibitem[\protect\citeauthoryear{Mohseni et~al.}{2022}]{Mohseni2022Ising}
\begin{barticle}
\bauthor{\bsnm{Mohseni}, \binits{N.}},
\bauthor{\bsnm{McMahon}, \binits{P.L.}},
\bauthor{\bsnm{Byrnes}, \binits{T.}}:
\batitle{Ising machines as hardware solvers of combinatorial optimization
  problems}.
\bjtitle{Nature Reviews Physics}
\bvolume{4}(\bissue{6}),
\bfpage{363}--\blpage{379}
(\byear{2022})
\end{barticle}
\endbibitem

\bibitem[\protect\citeauthoryear{{Goto Hayato} and {Endo
  Kotaro}}{2021}]{GotoHayato2021Highperformance}
\begin{botherref}
\oauthor{\bsnm{{Goto Hayato}}},
\oauthor{\bsnm{{Endo Kotaro}}}:
High-performance combinatorial optimization based on classical mechanics.
SCIENCE ADVANCES,
10
(2021)
\end{botherref}
\endbibitem

\bibitem[\protect\citeauthoryear{{Saavan Patel}
  et~al.}{2020}]{SaavanPatel2020Ising}
\begin{botherref}
\oauthor{\bsnm{{Saavan Patel}}},
\oauthor{\bsnm{{Lili Chen}}},
\oauthor{\bsnm{{Philip Canoza}}},
\oauthor{\bsnm{{Sayeef Salahuddin}}}:
Ising {{Model Optimization Problems}} on a {{FPGA Accelerated Restricted
  Boltzmann Machine}}.
arXiv
(2020)
\end{botherref}
\endbibitem

\bibitem[\protect\citeauthoryear{Auffinger et~al.}{2013}]{auffinger2013random}
\begin{barticle}
\bauthor{\bsnm{Auffinger}, \binits{A.}},
\bauthor{\bsnm{Arous}, \binits{G.B.}},
\bauthor{\bsnm{{\v C}ern{\`y}}, \binits{J.}}:
\batitle{Random matrices and complexity of spin glasses}.
\bjtitle{Communications on Pure and Applied Mathematics}
\bvolume{66}(\bissue{2}),
\bfpage{165}--\blpage{201}
(\byear{2013})
\end{barticle}
\endbibitem

\bibitem[\protect\citeauthoryear{Ros and
  Fyodorov}{2023}]{Ros2023highdimensional}
\begin{botherref}
\oauthor{\bsnm{Ros}, \binits{V.}},
\oauthor{\bsnm{Fyodorov}, \binits{Y.V.}}:
The high-dimensional landscape paradigm: {{Spin-glasses}}, and beyond.
Spin Glass Theory and Far Beyond: Replica Symmetry Breaking After 40 Years,
95--114
(2023)
\end{botherref}
\endbibitem

\bibitem[\protect\citeauthoryear{Billoire et~al.}{2018}]{billoire2018dynamic}
\begin{barticle}
\bauthor{\bsnm{Billoire}, \binits{A.}},
\bauthor{\bsnm{Fernandez}, \binits{{\relax LA}.}},
\bauthor{\bsnm{Maiorano}, \binits{A.}},
\bauthor{\bsnm{Marinari}, \binits{E.}},
\bauthor{\bsnm{{Martin-Mayor}}, \binits{V.}},
\bauthor{\bsnm{{Moreno-Gordo}}, \binits{J.}},
\bauthor{\bsnm{Parisi}, \binits{G.}},
\bauthor{\bsnm{{Ricci-Tersenghi}}, \binits{F.}},
\bauthor{\bsnm{{Ruiz-Lorenzo}}, \binits{J.}}:
\batitle{Dynamic variational study of chaos: Spin glasses in three dimensions}.
\bjtitle{Journal of Statistical Mechanics: Theory and Experiment}
\bvolume{2018}(\bissue{3}),
\bfpage{033302}
(\byear{2018})
\end{barticle}
\endbibitem

\bibitem[\protect\citeauthoryear{Franz et~al.}{2001}]{franz2001exact}
\begin{barticle}
\bauthor{\bsnm{Franz}, \binits{S.}},
\bauthor{\bsnm{Leone}, \binits{M.}},
\bauthor{\bsnm{{Ricci-Tersenghi}}, \binits{F.}},
\bauthor{\bsnm{Zecchina}, \binits{R.}}:
\batitle{Exact solutions for diluted spin glasses and optimization problems}.
\bjtitle{Physical review letters}
\bvolume{87}(\bissue{12}),
\bfpage{127209}
(\byear{2001})
\end{barticle}
\endbibitem

\bibitem[\protect\citeauthoryear{Kiss et~al.}{2024}]{kiss2024complete}
\begin{barticle}
\bauthor{\bsnm{Kiss}, \binits{A.}},
\bauthor{\bsnm{Zar{\'a}nd}, \binits{G.}},
\bauthor{\bsnm{Lovas}, \binits{I.}}:
\batitle{Complete replica solution for the transverse field
  {{Sherrington-Kirkpatrick}} spin glass model with continuous-time quantum
  {{Monte Carlo}} method}.
\bjtitle{Physical Review B}
\bvolume{109}(\bissue{2}),
\bfpage{024431}
(\byear{2024})
\end{barticle}
\endbibitem

\bibitem[\protect\citeauthoryear{Mo et~al.}{2023}]{mo2023nature}
\begin{barticle}
\bauthor{\bsnm{Mo}, \binits{J.}},
\bauthor{\bsnm{Chen}, \binits{H.}},
\bauthor{\bsnm{Chen}, \binits{R.}},
\bauthor{\bsnm{Jin}, \binits{M.}},
\bauthor{\bsnm{Liu}, \binits{M.}},
\bauthor{\bsnm{Xia}, \binits{Y.}}:
\batitle{Nature of spin-glass behavior of cobalt-doped iron disulfide
  nanospheres using the monte carlo method}.
\bjtitle{The Journal of Physical Chemistry C}
\bvolume{127}(\bissue{3}),
\bfpage{1475}--\blpage{1486}
(\byear{2023})
\end{barticle}
\endbibitem

\bibitem[\protect\citeauthoryear{Goto et~al.}{2019}]{goto2019combinatorial}
\begin{barticle}
\bauthor{\bsnm{Goto}, \binits{H.}},
\bauthor{\bsnm{Tatsumura}, \binits{K.}},
\bauthor{\bsnm{Dixon}, \binits{A.R.}}:
\batitle{Combinatorial optimization by simulating adiabatic bifurcations in
  nonlinear {{Hamiltonian}} systems}.
\bjtitle{Science advances}
\bvolume{5}(\bissue{4}),
\bfpage{2372}
(\byear{2019})
\end{barticle}
\endbibitem

\bibitem[\protect\citeauthoryear{{V Bapst} and {G
  Semerjian}}{2013}]{VBapst2013Thermal}
\begin{botherref}
\oauthor{\bsnm{{V Bapst}}},
\oauthor{\bsnm{{G Semerjian}}}:
Thermal, quantum and simulated quantum annealing: Analytical comparisons for
  simple models.
Journal of Physics,
9
(2013)
\end{botherref}
\endbibitem

\bibitem[\protect\citeauthoryear{{Francesco D'Angelo} and {Lucas
  B{\"o}ttcher}}{2020}]{FrancescoDAngelo2020Learning}
\begin{barticle}
\bauthor{\bsnm{{Francesco D'Angelo}}},
\bauthor{\bsnm{{Lucas B{\"o}ttcher}}}:
\batitle{Learning the {{Ising}} model with generative neural networks}.
\bjtitle{Physical Review Research}
\bvolume{2}(\bissue{2}),
\bfpage{023266}
(\byear{2020})
\doiurl{10.1103/PhysRevResearch.2.023266}
\end{barticle}
\endbibitem

\bibitem[\protect\citeauthoryear{Guo et~al.}{2019}]{Guo2019Attention}
\begin{barticle}
\bauthor{\bsnm{Guo}, \binits{S.}},
\bauthor{\bsnm{Lin}, \binits{Y.}},
\bauthor{\bsnm{Feng}, \binits{N.}},
\bauthor{\bsnm{Song}, \binits{C.}},
\bauthor{\bsnm{Wan}, \binits{H.}}:
\batitle{Attention {{Based Spatial-Temporal Graph Convolutional Networks}} for
  {{Traffic Flow Forecasting}}}.
\bjtitle{Proceedings of the AAAI Conference on Artificial Intelligence}
\bvolume{33}(\bissue{01}),
\bfpage{922}--\blpage{929}
(\byear{2019})
\doiurl{10.1609/aaai.v33i01.3301922} .
\bcomment{Chap. AAAI Technical Track: Applications}
\end{barticle}
\endbibitem

\bibitem[\protect\citeauthoryear{{S. Guo} et~al.}{2019}]{S.Guo2019Deep}
\begin{barticle}
\bauthor{\bsnm{{S. Guo}}},
\bauthor{\bsnm{{Y. Lin}}},
\bauthor{\bsnm{{S. Li}}},
\bauthor{\bsnm{{Z. Chen}}},
\bauthor{\bsnm{{H. Wan}}}:
\batitle{Deep {{Spatial}}--{{Temporal 3D Convolutional Neural Networks}} for
  {{Traffic Data Forecasting}}}.
\bjtitle{IEEE Transactions on Intelligent Transportation Systems}
\bvolume{20}(\bissue{10}),
\bfpage{3913}--\blpage{3926}
(\byear{2019})
\doiurl{10.1109/TITS.2019.2906365}
\end{barticle}
\endbibitem

\bibitem[\protect\citeauthoryear{Tang et~al.}{2019}]{Tang2019Traffic}
\begin{barticle}
\bauthor{\bsnm{Tang}, \binits{J.}},
\bauthor{\bsnm{Chen}, \binits{X.}},
\bauthor{\bsnm{Hu}, \binits{Z.}},
\bauthor{\bsnm{Zong}, \binits{F.}},
\bauthor{\bsnm{Han}, \binits{C.}},
\bauthor{\bsnm{Li}, \binits{L.}}:
\batitle{Traffic flow prediction based on combination of support vector machine
  and data denoising schemes}.
\bjtitle{Physica A: Statistical Mechanics and its Applications}
\bvolume{534},
\bfpage{120642}
(\byear{2019})
\end{barticle}
\endbibitem

\bibitem[\protect\citeauthoryear{Wu et~al.}{2018}]{Wu2018hybrid}
\begin{barticle}
\bauthor{\bsnm{Wu}, \binits{Y.}},
\bauthor{\bsnm{Tan}, \binits{H.}},
\bauthor{\bsnm{Qin}, \binits{L.}},
\bauthor{\bsnm{Ran}, \binits{B.}},
\bauthor{\bsnm{Jiang}, \binits{Z.}}:
\batitle{A hybrid deep learning based traffic flow prediction method and its
  understanding}.
\bjtitle{Transportation Research Part C: Emerging Technologies}
\bvolume{90},
\bfpage{166}--\blpage{180}
(\byear{2018})
\end{barticle}
\endbibitem

\bibitem[\protect\citeauthoryear{Xiao and Yin}{2019}]{Xiao2019Hybrid}
\begin{barticle}
\bauthor{\bsnm{Xiao}, \binits{Y.}},
\bauthor{\bsnm{Yin}, \binits{Y.}}:
\batitle{Hybrid {{LSTM}} neural network for short-term traffic flow
  prediction}.
\bjtitle{Information}
\bvolume{10}(\bissue{3}),
\bfpage{105}
(\byear{2019})
\end{barticle}
\endbibitem

\bibitem[\protect\citeauthoryear{Shaikh et~al.}{2022}]{Shaikh2022Review}
\begin{barticle}
\bauthor{\bsnm{Shaikh}, \binits{P.W.}},
\bauthor{\bsnm{{El-Abd}}, \binits{M.}},
\bauthor{\bsnm{Khanafer}, \binits{M.}},
\bauthor{\bsnm{Gao}, \binits{K.}}:
\batitle{A {{Review}} on {{Swarm Intelligence}} and {{Evolutionary Algorithms}}
  for {{Solving}} the {{Traffic Signal Control Problem}}}.
\bjtitle{IEEE TRANSACTIONS ON INTELLIGENT TRANSPORTATION SYSTEMS}
\bvolume{23}(\bissue{1}),
\bfpage{48}--\blpage{63}
(\byear{2022})
\doiurl{10.1109/TITS.2020.3014296}
\end{barticle}
\endbibitem

\bibitem[\protect\citeauthoryear{Suzuki et~al.}{2013}]{Suzuki2013Chaotic}
\begin{barticle}
\bauthor{\bsnm{Suzuki}, \binits{H.}},
\bauthor{\bsnm{Imura}, \binits{J.-i.}},
\bauthor{\bsnm{Aihara}, \binits{K.}}:
\batitle{Chaotic {{Ising-like}} dynamics in traffic signals}.
\bjtitle{Scientific Reports}
\bvolume{3}(\bissue{1}),
\bfpage{1127}
(\byear{2013})
\doiurl{10.1038/srep01127}
\end{barticle}
\endbibitem

\bibitem[\protect\citeauthoryear{Cook et~al.}{1994}]{Cook1994Combinatorial}
\begin{barticle}
\bauthor{\bsnm{Cook}, \binits{W.J.}},
\bauthor{\bsnm{Cunningham}, \binits{W.H.}},
\bauthor{\bsnm{Pulleyblank}, \binits{W.R.}},
\bauthor{\bsnm{Schrijver}, \binits{A.}}:
\batitle{Combinatorial optimization}.
\bjtitle{Unpublished manuscript}
\bvolume{10},
\bfpage{75}--\blpage{93}
(\byear{1994})
\end{barticle}
\endbibitem

\bibitem[\protect\citeauthoryear{Boik}{2008}]{Boik2008implicit}
\begin{barticle}
\bauthor{\bsnm{Boik}, \binits{R.J.}}:
\batitle{An implicit function approach to constrained optimization with
  applications to asymptotic expansions}.
\bjtitle{Journal of multivariate analysis}
\bvolume{99}(\bissue{3}),
\bfpage{465}--\blpage{489}
(\byear{2008})
\end{barticle}
\endbibitem

\bibitem[\protect\citeauthoryear{Bulos and Khuri}{2005}]{Bulos2005modified}
\begin{barticle}
\bauthor{\bsnm{Bulos}, \binits{B.}},
\bauthor{\bsnm{Khuri}, \binits{S.A.}}:
\batitle{On the modified {{Taylor}}'s approximation for the solution of linear
  and nonlinear equations}.
\bjtitle{Applied mathematics and computation}
\bvolume{160}(\bissue{3}),
\bfpage{939}--\blpage{953}
(\byear{2005})
\end{barticle}
\endbibitem

\bibitem[\protect\citeauthoryear{Mouroutsos and
  Sparis}{1985}]{Mouroutsos1985Taylor}
\begin{barticle}
\bauthor{\bsnm{Mouroutsos}, \binits{S.G.}},
\bauthor{\bsnm{Sparis}, \binits{P.D.}}:
\batitle{Taylor series approach to system identification, analysis and optimal
  control}.
\bjtitle{Journal of the Franklin Institute}
\bvolume{319}(\bissue{3}),
\bfpage{359}--\blpage{371}
(\byear{1985})
\end{barticle}
\endbibitem

\bibitem[\protect\citeauthoryear{Nourazar and
  {Nazari-Golshan}}{2015}]{Nourazar2015new}
\begin{barticle}
\bauthor{\bsnm{Nourazar}, \binits{S.S.}},
\bauthor{\bsnm{{Nazari-Golshan}}, \binits{A.}}:
\batitle{A new modification to homotopy perturbation method combined with
  {{Fourier}} transform for solving nonlinear {{Cauchy}} reaction diffusion
  equation}.
\bjtitle{Indian Journal of Physics}
\bvolume{89},
\bfpage{61}--\blpage{71}
(\byear{2015})
\end{barticle}
\endbibitem

\bibitem[\protect\citeauthoryear{Brush}{1967}]{Brush1967History}
\begin{barticle}
\bauthor{\bsnm{Brush}, \binits{S.G.}}:
\batitle{History of the {{Lenz-Ising}} model}.
\bjtitle{Reviews of modern physics}
\bvolume{39}(\bissue{4}),
\bfpage{883}
(\byear{1967})
\end{barticle}
\endbibitem

\bibitem[\protect\citeauthoryear{Cipra}{1987}]{Cipra1987introduction}
\begin{barticle}
\bauthor{\bsnm{Cipra}, \binits{B.A.}}:
\batitle{An introduction to the {{Ising}} model}.
\bjtitle{The American Mathematical Monthly}
\bvolume{94}(\bissue{10}),
\bfpage{937}--\blpage{959}
(\byear{1987})
\end{barticle}
\endbibitem

\bibitem[\protect\citeauthoryear{Aoki and Kobayashi}{2016}]{Aoki2016Restricted}
\begin{barticle}
\bauthor{\bsnm{Aoki}, \binits{K.-I.}},
\bauthor{\bsnm{Kobayashi}, \binits{T.}}:
\batitle{Restricted {{Boltzmann}} machines for the long range {{Ising}}
  models}.
\bjtitle{Modern Physics Letters B}
\bvolume{30}(\bissue{34}),
\bfpage{1650401}
(\byear{2016})
\end{barticle}
\endbibitem

\bibitem[\protect\citeauthoryear{Gu and Zhang}{2022}]{Gu2022Thermodynamics}
\begin{barticle}
\bauthor{\bsnm{Gu}, \binits{J.}},
\bauthor{\bsnm{Zhang}, \binits{K.}}:
\batitle{Thermodynamics of the {{Ising}} model encoded in restricted
  {{Boltzmann}} machines}.
\bjtitle{Entropy}
\bvolume{24}(\bissue{12}),
\bfpage{1701}
(\byear{2022})
\end{barticle}
\endbibitem

\bibitem[\protect\citeauthoryear{Yoshioka
  et~al.}{2019}]{Yoshioka2019Transforming}
\begin{barticle}
\bauthor{\bsnm{Yoshioka}, \binits{N.}},
\bauthor{\bsnm{Akagi}, \binits{Y.}},
\bauthor{\bsnm{Katsura}, \binits{H.}}:
\batitle{Transforming generalized {{Ising}} models into {{Boltzmann}}
  machines}.
\bjtitle{Physical Review E}
\bvolume{99}(\bissue{3}),
\bfpage{032113}
(\byear{2019})
\end{barticle}
\endbibitem

\bibitem[\protect\citeauthoryear{Bybee et~al.}{2023}]{Bybee2023Efficient}
\begin{barticle}
\bauthor{\bsnm{Bybee}, \binits{C.}},
\bauthor{\bsnm{Kleyko}, \binits{D.}},
\bauthor{\bsnm{Nikonov}, \binits{D.E.}},
\bauthor{\bsnm{Khosrowshahi}, \binits{A.}},
\bauthor{\bsnm{Olshausen}, \binits{B.A.}},
\bauthor{\bsnm{Sommer}, \binits{F.T.}}:
\batitle{Efficient optimization with higher-order {{Ising}} machines}.
\bjtitle{Nature Communications}
\bvolume{14}(\bissue{1}),
\bfpage{6033}
(\byear{2023})
\end{barticle}
\endbibitem

\bibitem[\protect\citeauthoryear{Dan et~al.}{2020}]{Dan2020Clustering}
\begin{botherref}
\oauthor{\bsnm{Dan}, \binits{A.}},
\oauthor{\bsnm{Shimizu}, \binits{R.}},
\oauthor{\bsnm{Nishikawa}, \binits{T.}},
\oauthor{\bsnm{Bian}, \binits{S.}},
\oauthor{\bsnm{Sato}, \binits{T.}}:
Clustering approach for solving traveling salesman problems via {{Ising}} model
  based solver.
2020 57th ACM/IEEE Design Automation Conference (DAC),
1--6
(2020)
\end{botherref}
\endbibitem

\bibitem[\protect\citeauthoryear{Zhang et~al.}{2022}]{Zhang2022review}
\begin{botherref}
\oauthor{\bsnm{Zhang}, \binits{T.}},
\oauthor{\bsnm{Tao}, \binits{Q.}},
\oauthor{\bsnm{Liu}, \binits{B.}},
\oauthor{\bsnm{Han}, \binits{J.}}:
A review of simulation algorithms of classical ising machines for combinatorial
  optimization.
2022 IEEE International Symposium on Circuits and Systems (ISCAS),
1877--1881
(2022)
\end{botherref}
\endbibitem

\bibitem[\protect\citeauthoryear{Yamaoka et~al.}{2015}]{Yamaoka201520kspin}
\begin{barticle}
\bauthor{\bsnm{Yamaoka}, \binits{M.}},
\bauthor{\bsnm{Yoshimura}, \binits{C.}},
\bauthor{\bsnm{Hayashi}, \binits{M.}},
\bauthor{\bsnm{Okuyama}, \binits{T.}},
\bauthor{\bsnm{Aoki}, \binits{H.}},
\bauthor{\bsnm{Mizuno}, \binits{H.}}:
\batitle{A 20k-spin {{Ising}} chip to solve combinatorial optimization problems
  with {{CMOS}} annealing}.
\bjtitle{IEEE Journal of Solid-State Circuits}
\bvolume{51}(\bissue{1}),
\bfpage{303}--\blpage{309}
(\byear{2015})
\end{barticle}
\endbibitem

\bibitem[\protect\citeauthoryear{Katzgraber et~al.}{2008}]{Katzgraber2008New}
\begin{barticle}
\bauthor{\bsnm{Katzgraber}, \binits{H.}},
\bauthor{\bsnm{Hartmann}, \binits{A.}},
\bauthor{\bsnm{Young}, \binits{A.}}:
\batitle{New insights from one-dimensional spin glasses}.
\bjtitle{Physics Procedia}
\bvolume{6},
\bfpage{35}--\blpage{45}
(\byear{2008})
\doiurl{10.1016/J.PHPRO.2010.09.026}
\end{barticle}
\endbibitem

\bibitem[\protect\citeauthoryear{Zhu et~al.}{2016}]{Zhu2016Brief}
\begin{bchapter}
\bauthor{\bsnm{Zhu}, \binits{L.}},
\bauthor{\bsnm{Ikeda}, \binits{K.}},
\bauthor{\bsnm{Pang}, \binits{P.}},
\bauthor{\bsnm{Zhang}, \binits{R.}},
\bauthor{\bsnm{Sarrafzadeh}, \binits{A.}}:
\bctitle{A {{Brief Review}} of {{Spin-Glass Applications}} in {{Unsupervised}}
  and {{Semi-supervised Learning}}}.
In: \beditor{\bsnm{Hirose}, \binits{A.}},
\beditor{\bsnm{Ozawa}, \binits{S.}},
\beditor{\bsnm{Doya}, \binits{K.}},
\beditor{\bsnm{Ikeda}, \binits{K.}},
\beditor{\bsnm{Lee}, \binits{M.}},
\beditor{\bsnm{Liu}, \binits{D.}} (eds.)
\bbtitle{Neural {{Information Processing}}},
pp. \bfpage{579}--\blpage{586}.
\bpublisher{Springer},
\blocation{Cham}
(\byear{2016})
\end{bchapter}
\endbibitem

\bibitem[\protect\citeauthoryear{Chowdhury}{2014}]{Chowdhury2014Spin}
\begin{bbook}
\bauthor{\bsnm{Chowdhury}, \binits{D.}}:
\bbtitle{Spin Glasses and Other Frustrated Systems}.
\bpublisher{Princeton University Press},
\blocation{Princeton, NJ}
(\byear{2014})
\end{bbook}
\endbibitem

\bibitem[\protect\citeauthoryear{Inoue}{2003}]{Inoue2003Image}
\begin{barticle}
\bauthor{\bsnm{Inoue}, \binits{J.}}:
\batitle{Image restoration and ising spin glasses in a transverse field}.
\bjtitle{Physica Scripta}
\bvolume{2003},
\bfpage{70}--\blpage{76}
(\byear{2003})
\doiurl{10.1238/Physica.Topical.106a00070}
\end{barticle}
\endbibitem

\bibitem[\protect\citeauthoryear{Lustig et~al.}{2007}]{Lustig2007Sparse}
\begin{botherref}
\oauthor{\bsnm{Lustig}, \binits{M.}},
\oauthor{\bsnm{Donoho}, \binits{D.}},
\oauthor{\bsnm{Pauly}, \binits{J.}}:
Sparse {{MRI}}: {{The}} application of compressed sensing for rapid {{MR}}
  imaging.
Magnetic Resonance in Medicine
\textbf{58}
(2007)
\doiurl{10.1002/mrm.21391}
\end{botherref}
\endbibitem

\bibitem[\protect\citeauthoryear{Rosen and Goodwin}{1993}]{Rosen1993Large}
\begin{botherref}
\oauthor{\bsnm{Rosen}, \binits{B.}},
\oauthor{\bsnm{Goodwin}, \binits{J.M.}}:
Large scale simulations of a spin glass image associative memory.
IEEE International Conference on Neural Networks,
908--9132
(1993)
\doiurl{10.1109/ICNN.1993.298678}
\end{botherref}
\endbibitem

\bibitem[\protect\citeauthoryear{Goodwin et~al.}{1988}]{Goodwin1988Exploration}
\begin{botherref}
\oauthor{\bsnm{Goodwin}, \binits{J.M.}},
\oauthor{\bsnm{Rosen}, \binits{B.}},
\oauthor{\bsnm{Vidal}, \binits{J.}}:
Exploration of learning in an associative magnetic processor.
IEEE 1988 International Conference on Neural Networks,
197--2042
(1988)
\doiurl{10.1109/ICNN.1988.23929}
\end{botherref}
\endbibitem

\bibitem[\protect\citeauthoryear{Metz and Peron}{2021}]{Metz2021Mean-field}
\begin{botherref}
\oauthor{\bsnm{Metz}, \binits{F.L.}},
\oauthor{\bsnm{Peron}, \binits{T.}}:
Mean-field theory of vector spin models on networks with arbitrary degree
  distributions.
Journal of Physics: Complexity
\textbf{3}
(2021)
\doiurl{10.1088/2632-072X/ac4bed}
\end{botherref}
\endbibitem

\bibitem[\protect\citeauthoryear{Friedrichs and
  Wolynes}{1989}]{Friedrichs1989Toward}
\begin{barticle}
\bauthor{\bsnm{Friedrichs}, \binits{M.}},
\bauthor{\bsnm{Wolynes}, \binits{P.}}:
\batitle{Toward protein tertiary structure recognition by means of associative
  memory hamiltonians}.
\bjtitle{Science}
\bvolume{246},
\bfpage{371}--\blpage{373}
(\byear{1989})
\doiurl{10.1126/science.246.4928.371}
\end{barticle}
\endbibitem

\bibitem[\protect\citeauthoryear{Goldstein et~al.}{1992}]{Goldstein1992Optimal}
\begin{barticle}
\bauthor{\bsnm{Goldstein}, \binits{R.}},
\bauthor{\bsnm{{Luthey-Schulten}}, \binits{Z.}},
\bauthor{\bsnm{Wolynes}, \binits{P.}}:
\batitle{Optimal protein-folding codes from spin-glass theory.}
\bjtitle{Proceedings of the National Academy of Sciences of the United States
  of America}
\bvolume{89 11},
\bfpage{4918}--\blpage{22}
(\byear{1992})
\doiurl{10.1073/PNAS.89.11.4918}
\end{barticle}
\endbibitem

\bibitem[\protect\citeauthoryear{Kirkpatrick and
  Thirumalai}{1987}]{kirkpatrick1987p}
\begin{barticle}
\bauthor{\bsnm{Kirkpatrick}, \binits{T.R.}},
\bauthor{\bsnm{Thirumalai}, \binits{D.}}:
\batitle{P-spin-interaction spin-glass models: {{Connections}} with the
  structural glass problem}.
\bjtitle{Physical Review B}
\bvolume{36}(\bissue{10}),
\bfpage{5388}
(\byear{1987})
\end{barticle}
\endbibitem

\bibitem[\protect\citeauthoryear{KIRKPATRICK
  et~al.}{1983}]{kirkpatrick1983optimization}
\begin{barticle}
\bauthor{\bsnm{KIRKPATRICK}, \binits{S.}},
\bauthor{\bsnm{GELATT}, \binits{{\relax CD}.}},
\bauthor{\bsnm{VECCHI}, \binits{{\relax MP}.}}:
\batitle{{{OPTIMIZATION BY SIMULATED ANNEALING}}}.
\bjtitle{SCIENCE}
\bvolume{220}(\bissue{4598}),
\bfpage{671}--\blpage{680}
(\byear{1983})
\doiurl{10.1126/science.220.4598.671}
\end{barticle}
\endbibitem

\bibitem[\protect\citeauthoryear{Delahaye et~al.}{2019}]{delahaye2019simulated}
\begin{bchapter}
\bauthor{\bsnm{Delahaye}, \binits{D.}},
\bauthor{\bsnm{Chaimatanan}, \binits{S.}},
\bauthor{\bsnm{Mongeau}, \binits{M.}}:
\bctitle{Simulated {{Annealing}}: {{From Basics}} to {{Applications}}}.
In: \beditor{\bsnm{Gendreau}, \binits{M.}},
\beditor{\bsnm{Potvin}, \binits{J.-Y.}} (eds.)
\bbtitle{Handbook of {{Metaheuristics}}}
vol. \bseriesno{272},
pp. \bfpage{1}--\blpage{35}.
\bpublisher{Springer},
\blocation{Cham}
(\byear{2019}).
\doiurl{10.1007/978-3-319-91086-4_1}
\end{bchapter}
\endbibitem

\end{thebibliography}
